\documentclass[aip,jcp,showpacs,nobibnotes,reprint,floatfix]{revtex4-1}
\usepackage{graphics,graphicx,amsfonts,amsmath,amsbsy,amssymb,color}
\usepackage{verbatim}  
\usepackage{psfrag}  
\usepackage{tabularx}
\usepackage{xmpmulti}
\usepackage{hyperref}
\usepackage{marginnote}
\usepackage{subfigure}
\usepackage{paracol}
\usepackage{xparse}
\usepackage{layouts}
\usepackage{afterpage}
\usepackage{braket}
\usepackage{paralist}
\usepackage{bm}
\usepackage{url}
\usepackage[normalem]{ulem}
\usepackage{color}
\usepackage[colorinlistoftodos]{todonotes}
\usepackage[american]{babel} 

\newcommand{\refeq}[1]{{Eq.~(\ref{#1})}}
\newcommand{\reffig}[1]{{Fig.~\ref{#1}}}

\hypersetup{hyperindex,breaklinks,colorlinks=true,linkcolor=blue,citecolor=blue,urlcolor=blue,filecolor=blue}

\begin{document}

\title{Accelerating convergence to the thermodynamic limit with twist angle selection applied to methods beyond many-body perturbation theory}
\author{Tina~N.~Mihm$^{(i),(ii)}$}
\author{William~Z.~Van Benschoten$^{(i),(ii)}$}
\author{James~J.~Shepherd$^{(i),(ii)}$}
\email{james-shepherd@uiowa.edu}
\address{$^{(i)}$ Department of Chemistry, University of Iowa \\ $^{(ii)}$ University of Iowa Informatics Initiative, University of Iowa}

\begin{abstract}
We recently developed a scheme to use low-cost calculations to find a single twist angle where the couple cluster doubles energy of a single calculation matches the twist-averaged coupled cluster doubles energy in a finite unit cell.
We used initiator full configuration interaction quantum Monte Carlo ($i$-FCIQMC) as an example of an exact method beyond coupled cluster doubles theory to show that this selected twist angle approach had comparable accuracy in methods beyond coupled cluster. 
Further, at least for small system sizes, we show that the same twist angle can also be found by comparing the energy directly (at the level of second-order Moller-Plesset theory) suggesting a route toward twist angle selection which requires minimal modification to existing codes which can perform twist averaging. 
\end{abstract}
\date{\today}
\maketitle

\section{Introduction}

Wavefunction or many-body perturbation theory methods are currently under active development for the treatment of solids with periodic boundary conditions.\cite{gruber_applying_2018, gruber_ab_2018, liao_communication:_2016, mihm_optimized_2019, azadi_efficient_2019, booth_towards_2013,ruggeri_correlation_2018, shepherd_many-body_2013, gruneis_explicitly_2013, mihm_advanced_2020, shepherd_communication:_2016, shepherd_convergence_2012, shepherd_coupled_2014, holzmann_finite-size_2011, sun_pyscf:_2018, sun_pyscf_2018, wang_excitons_2020, mcclain_gaussian-based_2017, mcclain_spectral_2016, lewis_ab_2019, booth_plane_2016, dornheim_ab_2016, harl_assessing_2010, lebegue_cohesive_2010, schimka_accurate_2010, harl_accurate_2009, gruneis_making_2009, riemelmoser_plane_2020, motta_towards_2017, gruneis_efficient_2015, gruneis_second-order_2010, gruneis_explicitly_2013, liao_communication:_2016,irmler_duality_2019} %
A major source of error in these calculations comes from the approximation made when an infinite solid is modeled using a Hamiltonian with a fixed particle number. \cite{drummond_finite-size_2008, lin_twist-averaged_2001, chiesa_finite-size_2006, liao_communication:_2016, mostaani_quantum_2015, pierleoni_coupled_2004, gruber_applying_2018, azadi_efficient_2019, mihm_advanced_2020, shepherd_many-body_2013, shepherd_communication:_2016, holzmann_finite-size_2011, dornheim_ab_2016, fraser_finite-size_1996, williamson_elimination_1997, gruneis_explicitly_2013}
These errors, termed finite size errors, impede the development of high accuracy methods particularly as they are commonly limited by computational cost scaling in the system sizes they can treat.

In the quantum Monte Carlo community, twist averaging is used as a way to reduce the finite size error of a single calculation of fixed size.\cite{drummond_finite-size_2008, lin_twist-averaged_2001, chiesa_finite-size_2006, zong_spin_2002, pierleoni_coupled_2004, holzmann_finite-size_2011, filippi_quantum_1999, mostaani_quantum_2015}
By taking offsets in the $k$-point grid and averaging the energy over them, finite size errors can be reduced. 
Any property can, in principle, be twist-averaged; however, one of the main uses of this tool to date has been to smooth total energy extrapolation curves by twist averaging the total energy. %
In general, the cost to perform such calculations scales linearly with the number of offsets. 
These offsets can be chosen as random points, on a grid, or through some selection procedure.~\cite{drummond_finite-size_2008, baldereschi_mean-value_1973, rajagopal_variational_1995, kent_finite-size_1999,azadi_efficient_2019} 

Twist-averaging as applied to stochastic methods can frequently benefit from simultaneous removal of finite size errors and stochastic errors. 
However, when it comes to deterministic methods such as coupled cluster theory, there are no stochastic errors in the original method before twist averaging. 
This means the cost increases linearly with each twist angle, as a new calculation needs to be performed.
Savings which come from symmetries can be employed, but even so, the cost is typically substantial (around 10 to 100 times the cost of one calculation).

In an attempt to address the scaling of twist averaging for deterministic calculations, we recently developed a twist angle selection scheme called ``connectivity twist averaging''
\cite{mihm_optimized_2019}
The connectivity is a second-order approximation to how the Hamiltonian is connected by non-zero matrix elements; it has the same cost as second-order Moller-Plesset (MP2) theory. 
Here, connectivity refers to the relationship between the virtual orbitals and the occupied orbitals in terms of momentum transfer vectors.
Close to the Fermi surface, these momenta are not sampled in a way that is highly dependent on the twist angle. 
As these are precisely the excitations that can have significant weight in the correlation energy, it was our sense that these connections were a dominant cause of finite size effects.  We found that choosing a twist angle with a connectivity close to the twist-averaged result also meant that the energy at this twist angle mirrored the twist-averaged energy.

Our original paper showed that connectivity twist averaging was effective for coupled cluster doubles (CCD) theory, leaving open the question of whether the inclusion of higher order correlation effects would change our conclusions. 
Here, we show that this twist angle selection scheme also works for methods beyond CCD, by using initiator full configuration interaction quantum Monte Carlo ($i$-FCIQMC),\cite{cleland_communications:_2010} a recently developed QMC method.\cite{blunt_communication_2018, shepherd_investigation_2012, ruggeri_correlation_2018, shepherd_sign_2014, blunt_hybrid_2019, ghanem_unbiasing_2019, schwarz_insights_2015, thomas_symmetry_2014, petras_fully_2019}
Twist-averaged $i$-FCIQMC calculations are compared with single calculations at the selected twist angle. 
We find similar agreement between connectivity twist-averaged (cTA) and twist-averaged (TA) energies for $i$-FCIQMC, demonstrating that our approach is applicable to methods which have a more thorough description of correlation than those considered with coupled cluster truncated at the doubles level. 
We also investigate an alternative means to select a single twist angle: finding the twist angle that best reproduces the twist-averaged energy at a lower-level method (here, MP2). 
This is similar in spirit to previous work by Needs and Foulkes~\cite{rajagopal_variational_1995, rajagopal_quantum_1994}, but differs in that we are considering the correlation energy. 
We find that, for the systems considered here, the energy matching approach works just as well as the connectivity approach.
This leaves open the possibility for other groups to select a twist angle without having to develop a new code to compute the connectivity.
Overall, we believe that this will broadens the utility of our approach to twist angle selection. 

\section{Methods}

\subsection{Coupled cluster on the uniform electron gas} 

We use the typical conventions for applying CCD to the UEG, found in our previous studies and summarized here.\cite{shepherd_many-body_2013, shepherd_range-separated_2014, shepherd_coupled_2014}
We model our solids using the uniform electron gas (UEG), which is widely used as a benchmark system for solid state methods development.
Here, the UEG is modeled as a three-dimensional cube with box length, $L$, and a density of $\Omega \equiv L^3 = \frac{4}{3} \pi r_s^3 N$, where $r_s$ is the Wigner-Seitz radius and $N$ is the number of electrons in the system.
We work entirely in reciprocal space (centered at the $\Gamma$-point), as this allows for discrete orbitals based on momentum quantum numbers, $k$, to make up the wavefunction. 
We used a plane wave basis set composed of $M$ orbitals, $\phi_j \equiv \phi_j({\bf r}, \sigma) = \sqrt{\frac{1}{\Omega}} \emph{e}^{\emph{i}{\bf k}_j \cdot {\bf r}}\delta_{\sigma_j, \sigma}$, all of which have a kinetic energy lower than a cutoff, $E_{cut, M} = \frac{1}{2}k_\mathrm{cut}^2$. 
In this formalism, the finite basis set is set as though the electron gas does not have an offset or twist angle.
Ewald interactions were used, per convention, and the electron repulsion integrals, $v_{ijab}$, are written:
\begin{equation}
 v_{ijab}=\sum_{{\bf G}}
 v_{\bf G}\delta_{{\bf k}_{i} - {\bf k}_{a}, {\bf G}} \delta_{{\bf k}_{j}- {\bf k}_{b},-{\bf G}} \delta_{\sigma_{i} ,\sigma_{a}}\delta_{\sigma_{j},\sigma_{b}}
\label{E_int_eq}
\end{equation}
where $v_{\bf G} \propto 4\pi / | {\bf G}|^2$.
Here, $i$ and $j$ represent indices of occupied orbitals and $a$ and $b$ represent indices of virtual orbitals. 
The form of the electron repulsion integrals guarantees that momentum is conserved during excitations according to ${\bf k}_i - {\bf k}_a = {\bf k}_b - {\bf k}_j = {\bf G}$, where ${\bf G}$ is the momentum transfer vector. 
All calculations are based on Hatree—Fock (HF), with the HF eigenvalues comprising kinetic and exchange energies. 
The Madelung constant, $v_M$, has been included in all calculations and reduces the total energy by $\frac{1}{2} N v_M$. 
In the thermodynamic limit, the particle number (and thus the box length) is infinitely large, meaning $v_M$ goes to zero.

Coupled cluster theory is a many-body perturbation theory where an exponential form for the wavefuction is assumed:
\begin{equation}
    \Psi = e^{\hat{T}}  \Psi_\mathrm{HF} %
\end{equation}
Here, the cluster operator $\hat{T}$ is the sum over all amplitudes (labelled $t$) and the $\Psi_\mathrm{HF}$ is the HF wavefunction. 
In this work, we only include singles and doubles amplitudes, defined by their excitation ranks from the HF reference. This method is called coupled cluster singles and doubles (CCSD). 
The CCSD wavefunction can be used in the Schr\"{o}dinger equation with the UEG Hamiltonian to obtain the total energy of the system, $E_\mathrm{total} = \langle \Psi_\mathrm{HF} | \hat{H} | e^{\hat{T}}  \Psi_\mathrm{HF} \rangle$.
The correlation energy is then obtained by subtracting off the HF energy, i.e.:
\begin{equation}
E_\mathrm{corr} = E_\mathrm{tot} - E_\mathrm{HF}
\label{correlation_E_from_Tot}
\end{equation}
and the correlation energy for CCSD can be written:
\begin{equation}
\label{CCSD_Corr_E}
E_\mathrm{corr} = \frac{1}{4} \sum_{ijab}  t_{ijab} \bar{v}_{ijab}
\end{equation}
Where $\bar{v}_{ijab}$ are the antisymmeterized repulsion integrals and $t_{ijab}$ are the doubles amplitudes (present in $\hat{T}$).
Finally, we note that in the UEG, the singles amplitudes are not included in this sum, as they are zero by momentum symmetry. Thus, in the UEG, CCSD is equivalent to coupled cluster doubles (CCD); we will call the method we are using CCD for the rest of this work.

\subsection{Full configuration interaction quantum Monte Carlo}
For our quantum Monte Carlo calculations, we work with full configuration interaction quantum Monte Carlo (FCIQMC)\cite{booth_fermion_2009} as implemented in the HANDE package.\cite{spencer_hande-qmc_2019}
Our description below follows notation given in our prior studies for consistency.\cite{shepherd_investigation_2012}
FCIQMC is a stochastic wavefunction-based method that uses a linear combination of Slater determinants to form the wavefunction.
Each Slater determinant represents a configuration with $N$ electrons and $M$ single-electron spin orbital basis functions:
\begin{equation}
    \label{Sum_slater_det}
    \Phi_0 = \sum_{\bf j} C_{\bf j}| D_{\bf j}\rangle,
\end{equation}
In the case of FCIQMC in the UEG, the basis set is composed of normalized plane waves:
\begin{equation}
    \label{norm_antisym_prod_plnwave}
    D_{\bf j} = {\mathcal{A}}[\phi_i({\bf x}_i)\phi_j({\bf x}_j)...\phi_p({\bf x}_N)]
\end{equation}
Where ${\bf j}$ is the unique index for each determinant, $\phi_j$ are the single particle plane wave orbitals, and $\mathcal{A}$ is the anti-symmetrization operator.
The Hamiltonian ($H$) can be written in this basis and then diagonalized to find the FCI eigenstates, which are the exact solutions to the Schr\"{o}dinger Equation for this basis. 

Alternatively, when substituted into the imaginary-time Schr\"{o}dinger Equation, \refeq{Sum_slater_det} produces a set of coupled equations based on the determinant coefficients: 
\begin{equation}
    \label{coupled_Ham_with_determinants}
    -\frac{\emph{d} C_{\bf j}}{\emph{d}\tau} = (H_{{\bf j}{\bf j}} - E_{\mathrm{HF}} - \emph{S}) C_{\bf j} + \sum_{{\bf i}\neq {\bf j}}H_{{\bf j}{\bf i}} C_{\bf i} %
\end{equation}
Here, $\tau$ is the imaginary time, and the sum over ${\bf j}$ is taken over the singly- and doubly-excited determinants from ${\bf i}$ (as $H$ is a two-particle object).
A population of $N_\mathrm{w}$ walkers are then introduced to stochastically sample the wavefunction through simulating the determinant coefficients using a finite-difference version of \refeq{coupled_Ham_with_determinants}. 
The walker population is controlled through using annihilation, death/cloning and spawning rules, which are algorithmic interpretations of the terms in this equation.\cite{booth_fermion_2009} 
In particular, the sum over ${\bf j}$ is sampled once per Monte Carlo cycle.

The shift, $S$, is used as a population control for the walkers, and is set up such that it is held constant until the walkers have reached a set population. 
Then, it is allowed to vary around the total energy to make sure the total walker population stays within a certain range and allows the calculation to reach convergence, which is reached when (1) the energy and walker population both reach a constant value on average (representing $\frac{\emph{d} C_{\bf j}}{\emph{d}\tau}=0$), and (2) the walker population is sufficiently large.
At convergence, $S$ can be averaged as an estimate of the ground-state energy. 
The estimate we use for this study is the projected energy: $\langle \Phi_0 | \Psi_\mathrm{HF} \rangle$, which agrees with $S$ for converged calculations but typically, here, has a smaller stochastic error.
For FCIQMC, then, this produces a set of master equations that can be used to measure the dynamics of the determinant coefficients in imaginary time for a given system.

The FCI space grows exponentially with the size of the system and the initiator approximation ($i$-FCIQMC) is used to help combat this scaling through further refining which walkers are used.\cite{cleland_communications:_2010, booth_breaking_2011, booth_approaching_2010, shepherd_investigation_2012, shepherd_full_2012, ghanem_unbiasing_2019} Here, the determinant space is divided up using a set constant, $n_\mathrm{add}$. If a determinant has a population greater than $n_\mathrm{add}$, walkers are allowed to spawn to empty determinants, and the determinant whose walkers are allowed to spawn is labeled an initiator determinant. %
This helps the calculation converge faster by refining the population and removing some of the stochastic error. Our prior work indicates that $n_\mathrm{add}$ above a certain number simply causes unnecessary duplication of walkers to resolve the same quality of wavefunction and $n_\mathrm{add}=3$ gives a good compromise.\cite{shepherd_investigation_2012}
It is important to note that in the large walker limit, $i$-FCIQMC converges to the FCI energy, as has been shown in previous work.\cite{cleland_communications:_2010,booth_breaking_2011, shepherd_investigation_2012} 

\subsection{Twist averaging} 

Twist averaging is used to reduce finite size effects and produce a smoother convergence to the thermodynamic limit.
The finite size errors that twist averaging overcomes are those associated with shell filling effects.\cite{drummond_finite-size_2008, lin_twist-averaged_2001} 
The shells here are made of degenerate orbitals. 
Thus, this form of averaging helps to reduce the random fluctuation at each $N$ seen in the $\Gamma$-point energy, providing a smoother extrapolation to the thermodynamic limit.   
Twist averaging addresses this issue by applying a small momentum offset, ${\bf k}_s$, to the orbitals:
\begin{equation} 
\phi_j \propto \exp(\overline{i}({\bf k}_j - {\bf k}_s) \cdot {\bf r}) 
\label{TA_orbs}
\end{equation} 
This offset breaks orbital degeneracy. 
We refer to the ${\bf k}_s$ as \emph{twist angles} for consistency with original work.%
\cite{lin_twist-averaged_2001}
The twist-averaged energy is then the average over all twist angles. For the correlation energy, this is:
\begin{equation}
\langle E_\mathrm{corr} \rangle_{{\bf k}_s} = \frac{1}{N_{\bf{k}_s}} \sum^{N_{\bf{k}_s}}_{t=1} E_\mathrm{corr}({\bf k}_{s,t})
\label{twist_Av_Ecorr}
\end{equation}
In this equation, ${N_{\bf{k}_s}}$ is the total number of twist angles used.
As a single calculation typically has to be run at each twist angle, the cost of twist averaging scales as ${N_{\bf{k}_s}}$.  
Here, we will address the method of twist averaging as applied to the correlation energy exclusively, in part because it is traditional in quantum chemistry to assume that it is best to separate the much easier/cheaper to converge HF component from the more expensive correlation component.

\subsection{Twist averaging in \emph{i}-FCIQMC}
Twist-averaging in $i$-FCIQMC follows similar conventions as in CCD, which were outlined above, with the exception that we used 50 random twist angles rather than 100 due to the higher cost of $i$-FCIQMC.  
To maintain the most consistency between our methods (CCD and FCIQMC) and with our prior work,\cite{mihm_optimized_2019} we use a random twist angle approach for our FCIQMC calculations even though we note other approaches in the literature.~\cite{ruggeri_correlation_2018}

Each $i$-FCIQMC calculation was run using the fixed shift strategy,\cite{shepherd_investigation_2012} where the shift is not varied throughout the calculation and the population is allowed to grow.
For each twist angle, one $i$-FCIQMC calculation was performed. 
Then, the different simulations were averaged.
This removes the stochastic error from $i$-FCIQMC and performs the twist averaging simultaneously.

The optimal shift for each $r_s$ value was found in previous work and, in general, results were insensitive to the choice of shift because initiator convergence was faster than expected.
Initiator error was removed by growing the population (increasing $N_w$) until the energy converges. 

\subsection{Connectivity Twist Averaging} 
As stated above, twist averaging has a linear cost scaling of $N_{{\bf k}_s}$. %
Over the years, there have been efforts in the QMC community to find the twist-averaged energy by using only a single twist angle\cite{rajagopal_quantum_1994, rajagopal_variational_1995, kent_finite-size_1999,drummond_finite-size_2008} or a small number of twist angles.\cite{azadi_efficient_2019}
These methods come with drawbacks, such as the Balderschi point\cite{baldereschi_mean-value_1973} being more effective for insulators (whether applied to $k$-points or twist angles).
Recently, we developed a method call connectivity twist averaging (cTA) that analyzes the way that the occupied space is connected to the virtual space through two-electron excitations. %
For a given excitation in the UEG consisting of two occupied orbitals ($i,j$) and two virtual orbitals ($a,b$), momentum conservation can zero the four-index integral ($v_{ijab}$).
The following equation expresses momentum conservation:
\begin{equation}
\eta_{ijab,{\bf G}}=\delta_{k_{i} - k_{a},{\bf G}} \delta_{k_{j}- k_{b},-{\bf G}}\delta_{\sigma_{i}\sigma_{a}}\delta_{\sigma_{j}\sigma_{b}}
\end{equation}
through setting $\eta_{ijab,{\bf G}}$ to 0 or 1 depending on whether the excitation is disallowed or allowed respectively.

What we term the connectivity is measured through a vector labelled ${\bf h}$\footnote{with reference to the idea of a histogram} with elements given by the expression:
\begin{equation}
h_x = \sum_{ijab} \eta_{ijab,G_x} +\sum_{ijab}\eta_{ijba,G_x} .
\label{cTA_hystogram}
\end{equation}
The integer index $x$ can be found as a function of $G$, $x = (G \frac{L}{2\pi})^2$.

Here, the form of the sum may be familiar as it is intended to be the same sum as the MP2 correlation energy.

A special twist angle is then selected by finding the single twist angle whose connectivity vector most closely matches the twist-averaged connectivity vector. 
Formally, we can define the sum of residuals as: 
\begin{equation}
S_\mathrm{res} ({\bf k}_s)=\sum_{x}\frac{1}{x^2}|h_x({\bf{k}_s}) - \langle h_x\rangle_{\bf{k}_s}|^2,
\label{cTA_min_sum}
\end{equation}
The special twist angle, $\bf{k}_s^*$, is calculated as the $\bf{k}_s$ that minimizes $S_\mathrm{res}({\bf k}_s)$.
The factor $1/x^2$ weights the sum towards shorter momentum vectors -- that are energetically more relevant -- which improves convergence at larger basis set sizes.

In our original work,\cite{mihm_optimized_2019} we found that CCD energy computed using the special twist angle $\bf{k}_s^*$ is very similar to the twist-averaged CCD energy. 
We found that it was useful to include twist-averaged HF eigenvalues (instead of the HF eigenvalues for a single twist angle) in addition to the twist angle selection, as this improved the agreement between the cTA and TA energies for small systems. 
This was impractical for this proof-of-concept $i$-FCIQMC study due to details of the implementation of the UEG in the code we are using, but is something that we will explore in the future. 
Thus, comparisons are made between CCD and $i$-FCIQMC that omit the averaging over eigenvalues.
For clarity, we will re-emphasize this in the text throughout.

Overall, as only one calculation needs to be performed, the cTA method reduces the cost of twist averaging by a factor of $ N_{{\bf k}_s}$. 
One of the advantages of this method in comparison to the Baldereschi point is that, instead of using one special twist angle for all systems with the same symmetry, it selects different twist angles depending on the system. We hope that this will allow for a broader applicability. %
In our previous study, we applied this method to CCD and showed good comparisons between TA-CCD and cTA-CCD. 
Here, we will expand our application of cTA to methods a more complete description of correlation, here FCIQMC, to show that these calculations can also benefit from the cTA method.

\section{Results}

\subsection{Initiator convergence in TA-\emph{i}-FCIQMC}

\begin{figure}
	\includegraphics[width=0.4\textwidth,height=\textheight,keepaspectratio]{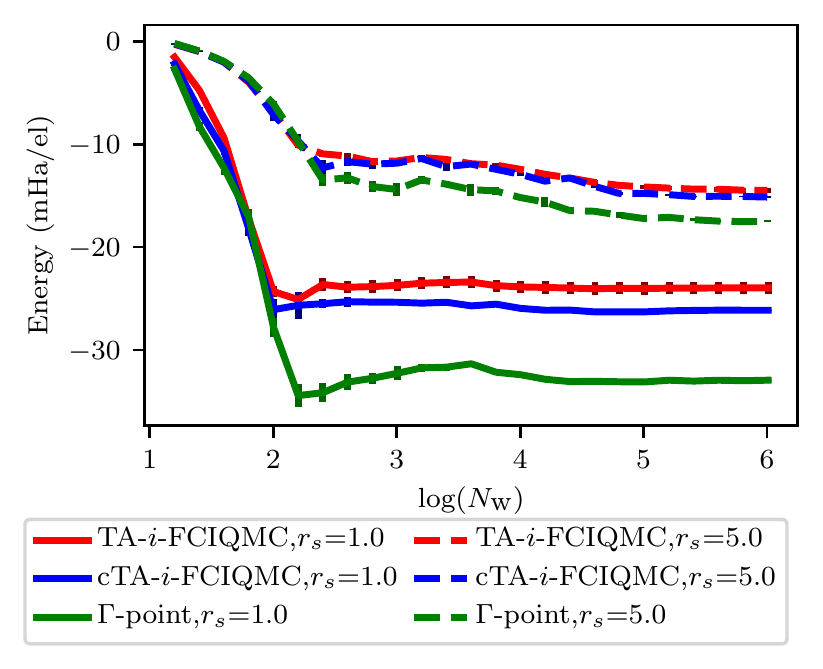}
	\caption{Comparison between connectivity twist averaging (special) and twist averaging (random) results are shown for two different densities for a UEG of $N=14, M=114$ The convergence of the energies for $i$-FCIQMC shows slower convergence for the larger $r_s$.}
	\label{subfig:1a}
\end{figure}

\begin{figure}
	\includegraphics[width=0.4\textwidth,height=\textheight,keepaspectratio]{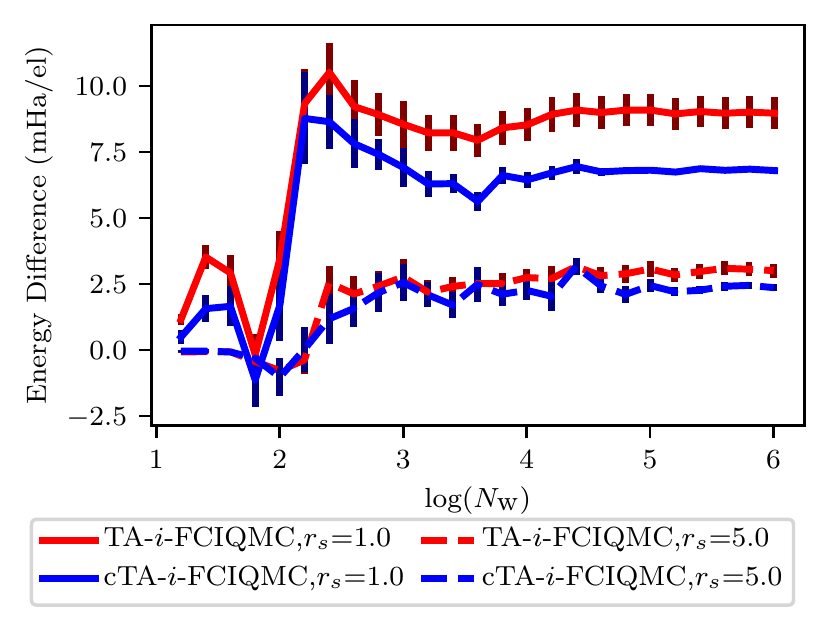}
	\caption{Comparison between connectivity twist averaging (special) and twist averaging (random) results are shown for two different densities for a UEG of $N=14, M=114$ The convergence of energy differences for $i$-FCIQMC is shown to be different than in \reffig{subfig:1a}.}
	\label{subfig:1c}
\end{figure}

The purpose of this paper is to use $i$-FCIQMC to compute high accuracy correlation energies that includes correlation beyond coupled cluster theory. %
Here, we choose a way to twist-average $i$-FCIQMC energies that is most consistent with how we used twist-averaging for coupled cluster theory in our original paper on the connectivity twist averaging approach. 
An important aspect of an $i$-FCIQMC calculation is the convergence of the energy with respect to the walker number.
One of the challenges with using $i$-FCIQMC is the tendency toward size inconsistency when not properly converged with walker number, which can result in oscillatory convergence of energy differences.
Many methods have been proposed to overcome the $i$-FCIQMC error.\cite{blunt_communication_2018} %
However, for the purposes of this study, we found that $i$-FCIQMC error was much more straightforward to converge with walker number.
To see this, we make comparison with the convergence data in work by Alavi and coworkers, who studied the $N=14$ uniform electron gas.\cite{shepherd_investigation_2012}
In so doing, they identified that as the density decreases, the number of walkers required to converge the energy increases. In particular, an $r_s$ of $5.0$ requires populations of about $10^9$ to obtain convergence. 

Figure \ref{subfig:1a} shows the $i$-FCIQMC energy as a function of walker number ($N_w$) for two different densities. 
The energy is said to be converged with initiator error in these graphs when increasing $N_w$ does not change the energy.
In \reffig{subfig:1a}, the convergence of the TA-$i$-FCIQMC and cTA-$i$-FCIQMC energies can be seen in comparison to the $\Gamma$-point energy convergence for $r_s  = 1.0$.
The $\Gamma$-point data match previously studied calculations by Alavi and coworkers, appearing to fully converge around $N_w=10^5$.
The twist-averaged energies show a somewhat faster convergence with $N_w$ but have a larger stochastic error, which may hide the convergence trend. 
The large error is mostly attributable to the difference in energy between different twist angles (i.e., not $i$-FCIQMC itself). 
Also in \reffig{subfig:1a}, this analysis is repeated for a different density of $r_s  = 5.0$. 
In contrast with $r_s  = 1.0$, the $r_s  = 5.0$ energies are somewhat slower to converge. 
This probably does not match the prior observations of taking $10^9$ walkers,\cite{shepherd_investigation_2012} but is still more significant than at $r_s=1.0$. 

We are particularly interested in energy differences to the $\Gamma$-point energy (i.e. $\Delta E_{\mathrm{TA}} = E_{\mathrm{TA}} - E_{\Gamma}$), because taking these differences smooths out the TDL extrapolation curve.%
Now, energy differences are converged at the same rate for both densities (\reffig{subfig:1c}). 
This fast convergence presumably results from error cancellation of two calculations with similar initiator error.
While it is known that initiator error is not size consistent, for the same system with a different symmetry, the initiator error appears to have cancelled. 
This is important and useful for the scope of our study, as it means we can study a key energy difference (the TA correction to the $\Gamma$-point) for a wider range of systems where the total energy is significantly harder to converge.

\subsection{cTA reliability in \emph{i}-FCIQMC}

\begin{figure}
\includegraphics[width=0.4\textwidth,height=\textheight,keepaspectratio]{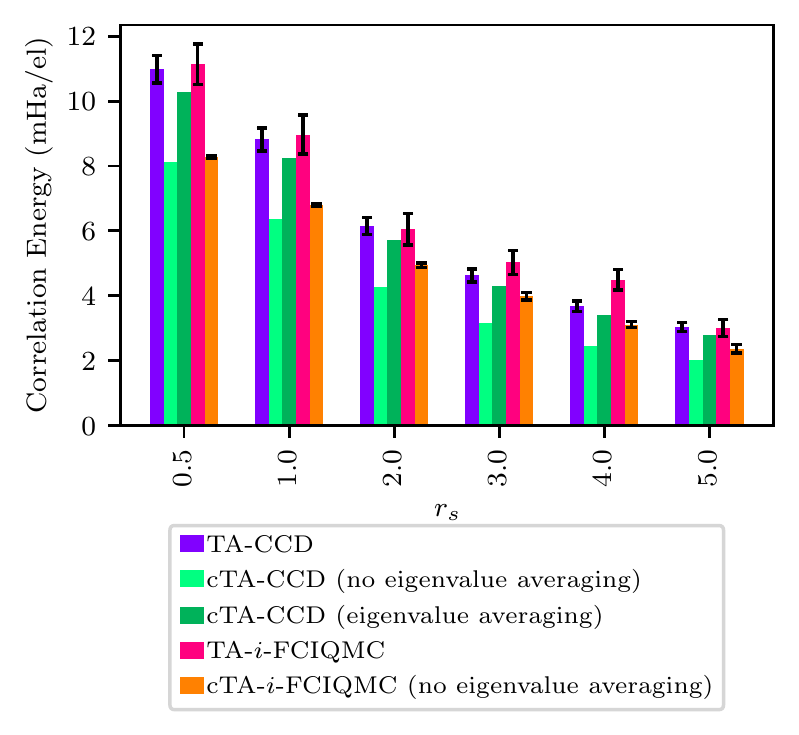}
\caption{Comparison across a range of densities between twist averaging and connectivity twist averaging for $i$-FCIQMC and CCD calculations on a $N=14$ system. All calculations are shown as a difference to the $\Gamma$-point. The cTA values are plotted with and without eigenvalue averaging (see text for discussion).}
\label{subfig:rs_data}
\end{figure}

\begin{figure}
\subfigure[\mbox{}]{%
\includegraphics[width=0.4\textwidth,height=\textheight,keepaspectratio]{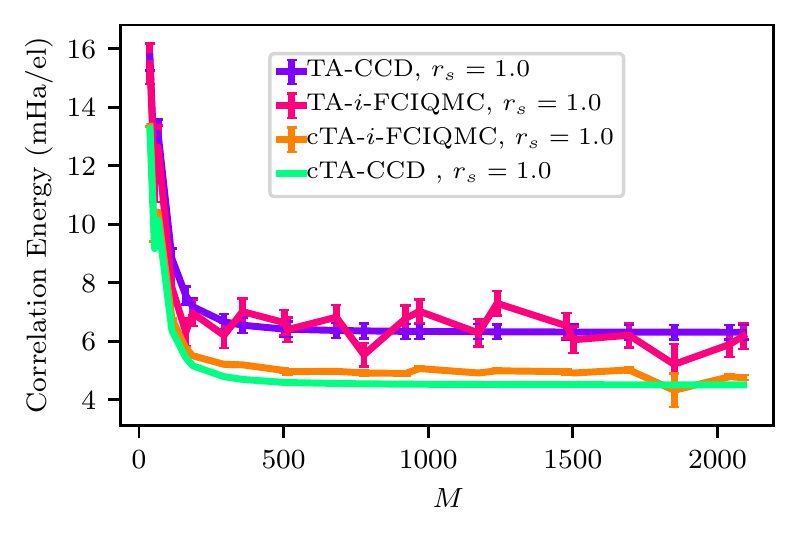}
\label{subfig:M_data_rs1}
}

\subfigure[\mbox{}]{%
\includegraphics[width=0.4\textwidth,height=\textheight,keepaspectratio]{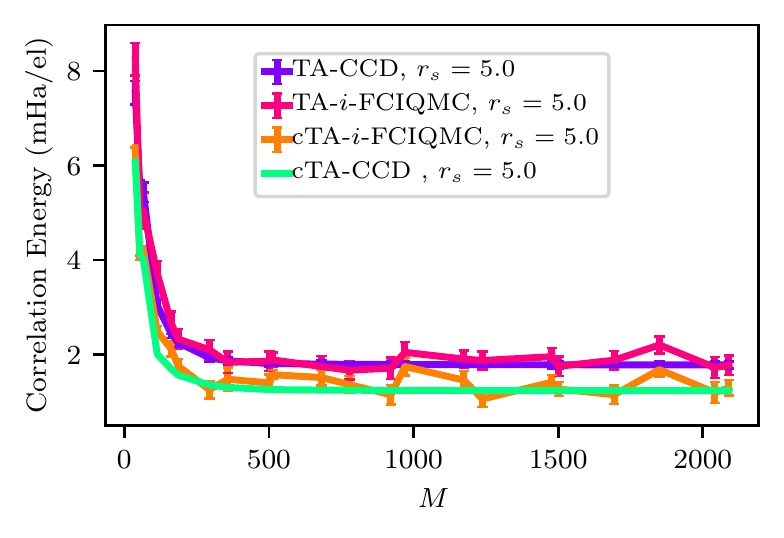}
\label{subfig:M_data_rs5}
}

\caption{Comparison between twist averaging and connectivity twist averaging for $i$-FCIQMC and CCD calculations on a $N=14$ system with varying basis sets at (a) $r_s=1.0$ and (b) $r_s=5.0$ . All energies are shown as a difference to the $\Gamma$-point. The cTA points are plotted without eigenvalue averaging.
} %
\label{fig:CCSD_methods_Comparision}
\end{figure}

\begin{figure}

\subfigure[\mbox{}]{%
\includegraphics[width=0.4\textwidth,height=\textheight,keepaspectratio]{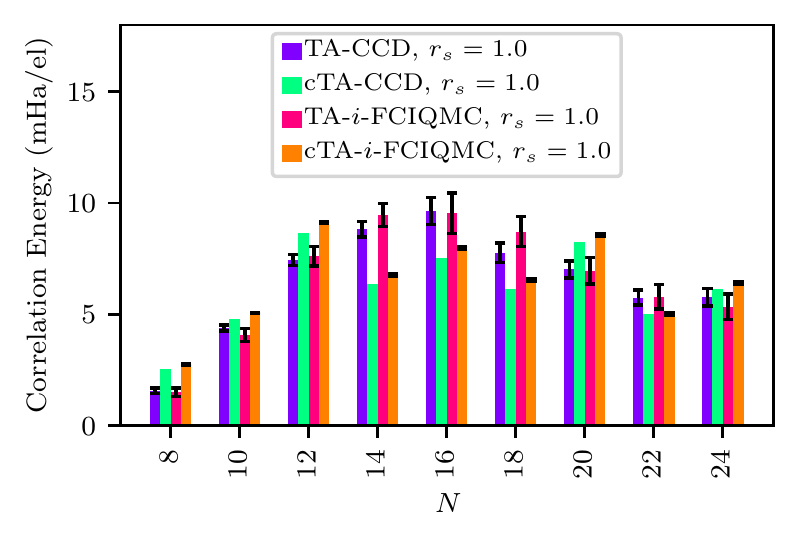}
\label{subfig:N_data_rs1}
}

\subfigure[\mbox{}]{%
\includegraphics[width=0.4\textwidth,height=\textheight,keepaspectratio]{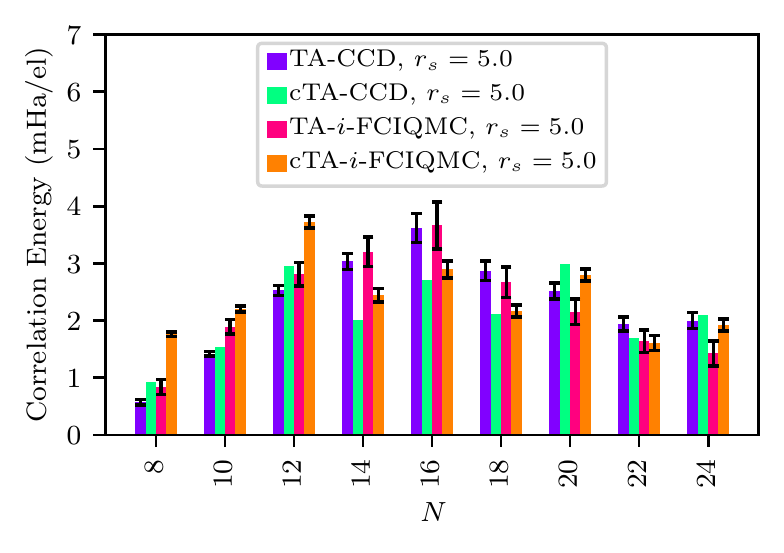}
\label{subfig:N_data_rs5}
}

\caption{Comparison between twist averaging and connectivity twist averaging for $i$-FCIQMC and CCD calculations on a $M=114$ system with varying electron numbers at (a) $r_s=1.0$ and (b) $r_s=5.0$ . All energies are shown as a difference to the $\Gamma$-point. The cTA points are plotted without eigenvalue averaging.}
\label{fig:N_data}
\end{figure}

\begin{figure}
\includegraphics[width=0.4\textwidth,height=\textheight,keepaspectratio]{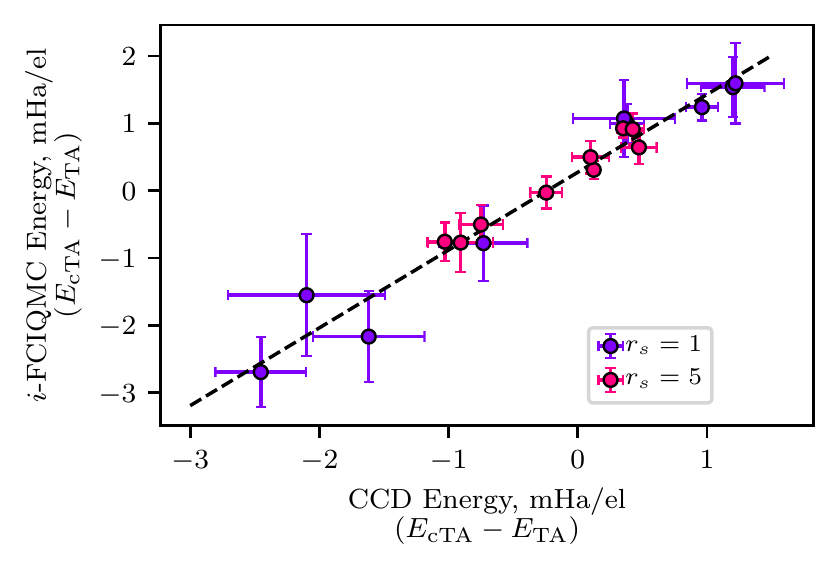}
\caption{The difference between twist averaging and connectivity twist averaging is plotted to make comparison between CCD and FCIQMC for the data set which varies electron number (previously shown in \reffig{fig:N_data}). The cTA points are plotted without eigenvalue averaging.}
\label{subfig:N_data_fcivsccd} %
\end{figure}

The purpose of this section is to address the question of whether methods that include correlation beyond CCD and more complex wavefunctions will similarly benefit from cTA. 
As in our previous paper, we will examine energy differences to the $\Gamma$-point (i.e. $\Delta E_{\mathrm{TA}} = E_{\mathrm{TA}} - E_{\Gamma}$) for the sake of clarity. 
Our goal is to relate these energy differences between CCSD and $i$-FCIQMC with a view to showing that cTA performs similarly to $i$-FCIQMC.
The main statistic we will use to show this is the mean absolute error. 

In \reffig{subfig:rs_data}, we show the results of cTA-$i$-FCIQMC compared with TA-$i$-FCIQMC for the $N=14$ electron gas at a range of $r_s$ values with a basis set of $M=114$.
Also shown in this figure is a comparison to CCD cTA and TA results, collected in a similar manner to $i$-FCIQMC.
For a given density ($r_s$), the same MP2 calculations were used, and so the cTA twist angle is the same for $i$-FCIQMC and CCD. 
The general trend of cTA-$i$-FCIQMC follows that of cTA-CCD in being significantly closer to the twist-averaged energy than the $\Gamma$-point whilst retaining a small amount of error.
The mean absolute errors (taken relative to TA) for this range of $r_s$ values is 1.8(2) mHa/electron for cTA-CCD and 1.5(5) mHa/electron for cTA-$i$-FCIQMC. %
Here, the numbers in parentheses represent the propagated stochastic error, mainly due to twist averaging.
We note one limitation is that these errors are substantial in absolute terms when using $i$-FCIQMC.
Small electron numbers like $N=14$ represent a particular challenge for a special twist angle method, as $N=14$ only has 4 unique correlation energies across all twist angles.\cite{lin_twist-averaged_2001} 
In our original work,\cite{mihm_optimized_2019} this was remedied by using eigenvalue averaging, which reduces the error for CCD to 0.4(2) and it could be expected that $i$-FCIQMC would behave similarly. 
These data form the first of our demonstrations that methods that include more correlation and parameters than CCD still benefit from the cTA approach.

In order to investigate how cTA-$i$-FCIQMC varies with basis set size, we computed cTA-$i$-FCIQMC and TA-$i$-FCIQMC for a range of basis sets (from $M = 38$ to $M = 2090$) for a set system size of $N=14$.
In Fig. \ref{subfig:M_data_rs1} and \ref{subfig:M_data_rs5}, we show the results for $r_s=1.0$ and $r_s=5.0$, respectively.
The mean absolute errors (compared with TA) for these two data sets are 1.3(2) mHa/electron for cTA-CCD and 1.1(4) mHa/electron for cTA-$i$-FCIQMC.
These agree within error, indicating that the performance of cTA is the same for both methods. 

In \reffig{subfig:N_data_rs1} and \reffig{subfig:N_data_rs5}, we show the results of cTA-$i$-FCIQMC compared with TA-$i$-FCIQMC for a range of system sizes from $N = 8$ to $N = 10$ for a set basis set of $M=114$. 
Following the trend of the other figures, the CCD, cTA, and TA comparison is also shown here with good agreement across system size, confirming the assertion of cTA-$i$-FCIQMC and cTA-CCD are of comparable quality in reproducing their respective twist-averaged energy.
The mean absolute errors for the electron number ranges (across both $r_s$ values) are 0.9(2) mHa/electron for cTA-CCD and 1.1(4) mHa/electron for cTA-$i$-FCIQMC

In \reffig{subfig:N_data_fcivsccd}, the $i$-FCIQMC results have been plotted as a function of CCD energies for the difference between the cTA and TA results for $N = 8 – 24$, $M = 114$ and $r_s = 1.0$ and $5.0$.  
Here, we see a linear relationship between the two methods across densities. 
This relationship shows that the cTA-$i$-FCIQMC error is almost identical to the cTA-CCD error. 

Overall, from these analyses, we find that the cTA method is transferable to FCIQMC (and, thus, other methods beyond CCD) based on its ability to obtain TA-$i$-FCIQMC energies with a similar accuracy to cTA-CCD. 
This is consistent with our hypothesis that the cTA method is transferable to total energies of different methods beyond many-body perturbation theory.

\subsection{Energy matching}

In our Communication on this method,\cite{mihm_optimized_2019} we had ruled out using the MP2 correlation energy as a way to perform twist angle selection because we were 
concerned about energy divergences for a metallic system. 
We had also preferred a method that would relate to a property of the wavefunction (which we termed the connectivity%
) over an energy measure.
However, since we now wish to make application to methods beyond CCD, it becomes more important to consider alternative ways to find a special twist angle. 
Here, we examine whether it is feasible to use MP2 as a way to select twist angles here. 

To use the MP2 energy, we introduce an energy matching criterion as a way to find the special twist angle.
In this variant of the method, $N_{\bf{k}_s}$ MP2 calculations at random twist angles are run. %
The special twist angle $\bf{k}_s^*$ is identified as the twist angle with the energy that is the minimum absolute difference to the twist-averaged energy, i.e., the twist angle that minimizes:
\begin{equation}
    S_\mathrm{res,emTA}=
    |E_{\mathrm{MP2}}({\bf{k}}_s) - \langle E_{\mathrm{MP2}} \rangle_{\bf{k}_s} |
\label{special_twist_EM}
\end{equation}
The twist angle selected in this way is used to calculate the CCD energy.
We refer to this procedure as energy matching twist averaging (emTA).

\begin{figure}
\includegraphics[width=0.45\textwidth,height=\textheight,keepaspectratio]{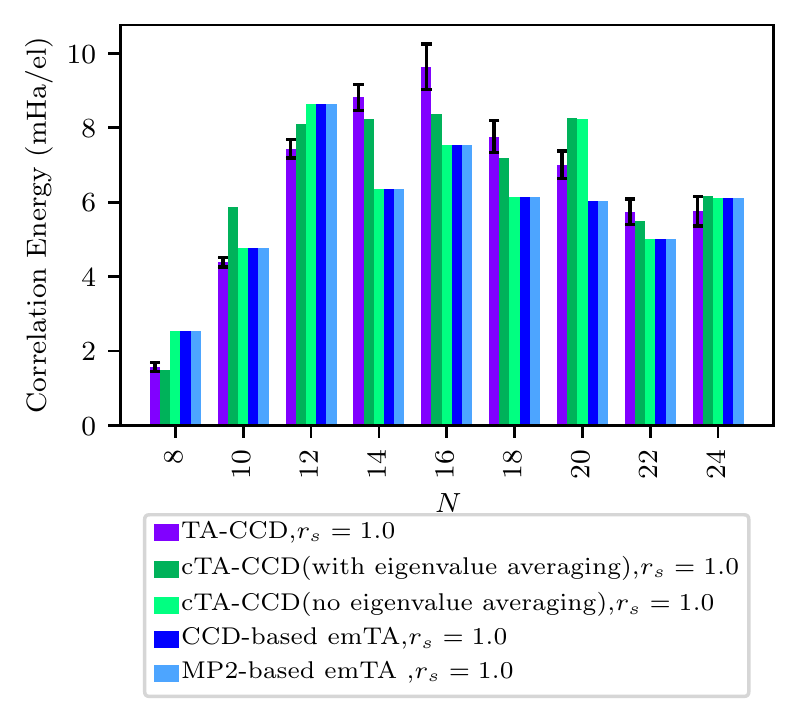}
\label{subfig:rs1_Nrange_CCSD}

\caption{Comparisons between twist averaging (TA) methods for CCD for a range of $N$ at $r_s = 1.0$. 
Energy matching (described in the text) is shown as emTA; connectivity twist averaging as cTA, which is calculated with eigenvalue averaging. 
As can be seen from the graph, cTA-CCD (without eigenvalue averaging) and emTA produce identical results except at $N=20$. All energies are shown as a difference to the $\Gamma$-point for clarity.
}
\label{fig:CCSD_methods_Comparison}
\end{figure}

In \reffig{fig:CCSD_methods_Comparison}, we show a comparison between the five twist averaging coupled cluster methods for a range of electron numbers, $N = 8 – 24$, at $r_s = 1.0$. 
The energies have all been graphed as a difference to the $\Gamma$-point as before. 
We find that the cTA energy closely resembles the emTA energy, i.e., cTA chooses the same twist angle as emTA over this range of systems. 
Exceptions to this are unusual; the only one we found was at $N=20$. 
We also looked at what happened when we replaced the MP2 energy with the CCD energy (denoted CCD-based emTA in \reffig{fig:CCSD_methods_Comparison}). 
This, too, produces the same energy. 
The mean absolute error (to TA) is $1.2(3)$ for emTA. %
It is important to note that this method is highly dependent on the quality of the MP2 energies and may not be a viable method at larger system sizes, where the MP2 energy diverges. 
The success of energy matching comes from the mean value theorem which, here, guarantees that the twist-averaged (mean) energy appears within the distribution of twist angles. 

\section{Discussion and concluding remarks}
The purpose of twist angle selection schemes is to accurately reproduce twist-averaged results from a single, special twist angle.
Connectivity twist averaging is one such selection scheme that uses an MP2-cost calculation for the selection process.
In our first paper, we showed this was effective for the CCD energy; here, we generalized this result to methods without a truncation by using $i$-FCIQMC. 
We performed $i$-FCIQMC calculations over a range of system sizes, basis sets, and densities for the uniform electron gas.
We were able to take advantage of a cancellation of error in the initiator error in FCIQMC, allowing us to take energy differences between calculations with different boundary conditions (i.e., periodic vs. twist-averaged), which converged quicker with walker number than the corresponding correlation energies. 
For the systems considered, we showed that the energy of the selected twist angle and the twist-averaged energy differed by a similar amount for $i$-FCIQMC when compared to our original CCD method. 
We then studied another twist angle selection scheme that we had set aside in our original work, which selects the twist angle by comparison of the MP2 energy. 
For these small systems, where the MP2 energy does not obviously diverge, the twist angle selection agreed between energy matching and connectivity twist averaging. 

There are a wide range of proposals in the literature for how to remedy finite size errors with no single clearly-established protocol. 
Some are obviously compatible with twist-averaging. Two classes of examples include extrapolations, which benefit from reduced extrapolation error~\cite{gruber_applying_2018, drummond_finite-size_2008, lin_twist-averaged_2001, chiesa_finite-size_2006, ruggeri_correlation_2018, gruber_applying_2018}; or correction schemes, where the correction is either additive or receives an additional benefit from a more balanced description of different electron numbers.~\cite{fraser_finite-size_1996, williamson_elimination_1997, kent_finite-size_1999, chiesa_finite-size_2006, kwee_finite-size_2008, drummond_finite-size_2008, dornheim_ab_2016,  holzmann_theory_2016, brown_path-integral_2013, liao_communication:_2016, gruneis_explicitly_2013} 
It is unclear or unlikely that twist averaging would be of use for other methods that reduce finite-size effects, including 
embedding theories such as DMET~\cite{bulik_electron_2014, knizia_density_2012}, DMFT~\cite{zhu_efficient_2020, choi_first-principles_2016}, SEET~\cite{zgid_finite_2017, iskakov_ab_2020} and density functional embedding theory~\cite{goodpaster_accurate_2014}; or DFT-based corrections.\cite{chiesa_finite-size_2006}%
Addressing the applicability of twist averaging to these methods is an open question beyond the scope of this paper.

We close noting two key limitations of our study.
The first is that we have not yet generalized connectivity twist averaging to real systems. This will be considered in a forthcoming separate paper. %
The second is that we have only studied smaller system sizes, a limitation imposed by $i$-FCIQMC. 
Despite this latter limitation, we still believe that our method is transferable to larger system sizes in-line with how CCD behaves. 

\section{Acknowledgements}
We gratefully acknowledge the University of Iowa for funding and computer resources through the University of Iowa Informatics Initiative. 
Code used throughout this was a locally modified version of a github repository used in previous work http://github.com/jamesjshepherd/uegccd ~\cite{shepherd_range-separated_2014,shepherd_coupled_2014}, and the HANDE-QMC package version 1.4 http://github.com/hande-qmc/hande ~\cite{spencer_hande-qmc_2019,spencer_open-source_2015}.
For the purposes of providing information about input options for the calculations used, files will be deposited with Iowa Research Online (IRO) with a reference number [to be inserted].
 
\section{Data Availability}
The data that supports the findings of this study are available within the article.


\begin{thebibliography}{72}%
\makeatletter
\providecommand \@ifxundefined [1]{%
 \@ifx{#1\undefined}
}%
\providecommand \@ifnum [1]{%
 \ifnum #1\expandafter \@firstoftwo
 \else \expandafter \@secondoftwo
 \fi
}%
\providecommand \@ifx [1]{%
 \ifx #1\expandafter \@firstoftwo
 \else \expandafter \@secondoftwo
 \fi
}%
\providecommand \natexlab [1]{#1}%
\providecommand \enquote  [1]{``#1''}%
\providecommand \bibnamefont  [1]{#1}%
\providecommand \bibfnamefont [1]{#1}%
\providecommand \citenamefont [1]{#1}%
\providecommand \href@noop [0]{\@secondoftwo}%
\providecommand \href [0]{\begingroup \@sanitize@url \@href}%
\providecommand \@href[1]{\@@startlink{#1}\@@href}%
\providecommand \@@href[1]{\endgroup#1\@@endlink}%
\providecommand \@sanitize@url [0]{\catcode `\\12\catcode `\$12\catcode
  `\&12\catcode `\#12\catcode `\^12\catcode `\_12\catcode `\%12\relax}%
\providecommand \@@startlink[1]{}%
\providecommand \@@endlink[0]{}%
\providecommand \url  [0]{\begingroup\@sanitize@url \@url }%
\providecommand \@url [1]{\endgroup\@href {#1}{\urlprefix }}%
\providecommand \urlprefix  [0]{URL }%
\providecommand \Eprint [0]{\href }%
\providecommand \doibase [0]{http://dx.doi.org/}%
\providecommand \selectlanguage [0]{\@gobble}%
\providecommand \bibinfo  [0]{\@secondoftwo}%
\providecommand \bibfield  [0]{\@secondoftwo}%
\providecommand \translation [1]{[#1]}%
\providecommand \BibitemOpen [0]{}%
\providecommand \bibitemStop [0]{}%
\providecommand \bibitemNoStop [0]{.\EOS\space}%
\providecommand \EOS [0]{\spacefactor3000\relax}%
\providecommand \BibitemShut  [1]{\csname bibitem#1\endcsname}%
\let\auto@bib@innerbib\@empty
%</preamble>
\bibitem [{\citenamefont {Gruber}\ \emph {et~al.}(2018)\citenamefont {Gruber},
  \citenamefont {Liao}, \citenamefont {Tsatsoulis}, \citenamefont {Hummel},\
  and\ \citenamefont {Grüneis}}]{gruber_applying_2018}%
  \BibitemOpen
  \bibfield  {author} {\bibinfo {author} {\bibfnamefont {T.}~\bibnamefont
  {Gruber}}, \bibinfo {author} {\bibfnamefont {K.}~\bibnamefont {Liao}},
  \bibinfo {author} {\bibfnamefont {T.}~\bibnamefont {Tsatsoulis}}, \bibinfo
  {author} {\bibfnamefont {F.}~\bibnamefont {Hummel}}, \ and\ \bibinfo {author}
  {\bibfnamefont {A.}~\bibnamefont {Grüneis}},\ }\href {\doibase
  10.1103/PhysRevX.8.021043} {\bibfield  {journal} {\bibinfo  {journal}
  {Physical Review X}\ }\textbf {\bibinfo {volume} {8}},\ \bibinfo {pages}
  {021043} (\bibinfo {year} {2018})}\BibitemShut {NoStop}%
\bibitem [{\citenamefont {Gruber}\ and\ \citenamefont
  {Grüneis}(2018)}]{gruber_ab_2018}%
  \BibitemOpen
  \bibfield  {author} {\bibinfo {author} {\bibfnamefont {T.}~\bibnamefont
  {Gruber}}\ and\ \bibinfo {author} {\bibfnamefont {A.}~\bibnamefont
  {Grüneis}},\ }\href {\doibase 10.1103/PhysRevB.98.134108} {\bibfield
  {journal} {\bibinfo  {journal} {Physical Review B}\ }\textbf {\bibinfo
  {volume} {98}},\ \bibinfo {pages} {134108} (\bibinfo {year}
  {2018})}\BibitemShut {NoStop}%
\bibitem [{\citenamefont {Liao}\ and\ \citenamefont
  {Grüneis}(2016)}]{liao_communication:_2016}%
  \BibitemOpen
  \bibfield  {author} {\bibinfo {author} {\bibfnamefont {K.}~\bibnamefont
  {Liao}}\ and\ \bibinfo {author} {\bibfnamefont {A.}~\bibnamefont
  {Grüneis}},\ }\href {\doibase 10.1063/1.4964307} {\bibfield  {journal}
  {\bibinfo  {journal} {The Journal of Chemical Physics}\ }\textbf {\bibinfo
  {volume} {145}},\ \bibinfo {pages} {141102} (\bibinfo {year}
  {2016})}\BibitemShut {NoStop}%
\bibitem [{\citenamefont {Mihm}, \citenamefont {McIsaac},\ and\ \citenamefont
  {Shepherd}(2019)}]{mihm_optimized_2019}%
  \BibitemOpen
  \bibfield  {author} {\bibinfo {author} {\bibfnamefont {T.~N.}\ \bibnamefont
  {Mihm}}, \bibinfo {author} {\bibfnamefont {A.~R.}\ \bibnamefont {McIsaac}}, \
  and\ \bibinfo {author} {\bibfnamefont {J.~J.}\ \bibnamefont {Shepherd}},\
  }\href {\doibase 10.1063/1.5091445} {\bibfield  {journal} {\bibinfo
  {journal} {The Journal of Chemical Physics}\ }\textbf {\bibinfo {volume}
  {150}},\ \bibinfo {pages} {191101} (\bibinfo {year} {2019})}\BibitemShut
  {NoStop}%
\bibitem [{\citenamefont {Azadi}\ and\ \citenamefont
  {Foulkes}(2019)}]{azadi_efficient_2019}%
  \BibitemOpen
  \bibfield  {author} {\bibinfo {author} {\bibfnamefont {S.}~\bibnamefont
  {Azadi}}\ and\ \bibinfo {author} {\bibfnamefont {W.~M.~C.}\ \bibnamefont
  {Foulkes}},\ }\href {\doibase 10.1103/PhysRevB.100.245142} {\bibfield
  {journal} {\bibinfo  {journal} {Physical Review B}\ }\textbf {\bibinfo
  {volume} {100}},\ \bibinfo {pages} {245142} (\bibinfo {year} {2019})},\
  \bibinfo {note} {publisher: American Physical Society}\BibitemShut {NoStop}%
\bibitem [{\citenamefont {Booth}\ \emph {et~al.}(2013)\citenamefont {Booth},
  \citenamefont {Grüneis}, \citenamefont {Kresse},\ and\ \citenamefont
  {Alavi}}]{booth_towards_2013}%
  \BibitemOpen
  \bibfield  {author} {\bibinfo {author} {\bibfnamefont {G.~H.}\ \bibnamefont
  {Booth}}, \bibinfo {author} {\bibfnamefont {A.}~\bibnamefont {Grüneis}},
  \bibinfo {author} {\bibfnamefont {G.}~\bibnamefont {Kresse}}, \ and\ \bibinfo
  {author} {\bibfnamefont {A.}~\bibnamefont {Alavi}},\ }\href {\doibase
  10.1038/nature11770} {\bibfield  {journal} {\bibinfo  {journal} {Nature}\
  }\textbf {\bibinfo {volume} {493}},\ \bibinfo {pages} {365} (\bibinfo {year}
  {2013})}\BibitemShut {NoStop}%
\bibitem [{\citenamefont {Ruggeri}, \citenamefont {Ríos},\ and\ \citenamefont
  {Alavi}(2018)}]{ruggeri_correlation_2018}%
  \BibitemOpen
  \bibfield  {author} {\bibinfo {author} {\bibfnamefont {M.}~\bibnamefont
  {Ruggeri}}, \bibinfo {author} {\bibfnamefont {P.~L.}\ \bibnamefont {Ríos}},
  \ and\ \bibinfo {author} {\bibfnamefont {A.}~\bibnamefont {Alavi}},\ }\href
  {\doibase 10.1103/PhysRevB.98.161105} {\bibfield  {journal} {\bibinfo
  {journal} {Physical Review B}\ }\textbf {\bibinfo {volume} {98}},\ \bibinfo
  {pages} {161105} (\bibinfo {year} {2018})}\BibitemShut {NoStop}%
\bibitem [{\citenamefont {Shepherd}\ and\ \citenamefont
  {Grüneis}(2013)}]{shepherd_many-body_2013}%
  \BibitemOpen
  \bibfield  {author} {\bibinfo {author} {\bibfnamefont {J.~J.}\ \bibnamefont
  {Shepherd}}\ and\ \bibinfo {author} {\bibfnamefont {A.}~\bibnamefont
  {Grüneis}},\ }\href {\doibase 10.1103/PhysRevLett.110.226401} {\bibfield
  {journal} {\bibinfo  {journal} {Physical Review Letters}\ }\textbf {\bibinfo
  {volume} {110}},\ \bibinfo {pages} {226401} (\bibinfo {year}
  {2013})}\BibitemShut {NoStop}%
\bibitem [{\citenamefont {Grüneis}\ \emph {et~al.}(2013)\citenamefont
  {Grüneis}, \citenamefont {Shepherd}, \citenamefont {Alavi}, \citenamefont
  {Tew},\ and\ \citenamefont {Booth}}]{gruneis_explicitly_2013}%
  \BibitemOpen
  \bibfield  {author} {\bibinfo {author} {\bibfnamefont {A.}~\bibnamefont
  {Grüneis}}, \bibinfo {author} {\bibfnamefont {J.~J.}\ \bibnamefont
  {Shepherd}}, \bibinfo {author} {\bibfnamefont {A.}~\bibnamefont {Alavi}},
  \bibinfo {author} {\bibfnamefont {D.~P.}\ \bibnamefont {Tew}}, \ and\
  \bibinfo {author} {\bibfnamefont {G.~H.}\ \bibnamefont {Booth}},\ }\href
  {\doibase doi:10.1063/1.4818753} {\bibfield  {journal} {\bibinfo  {journal}
  {The Journal of Chemical Physics}\ }\textbf {\bibinfo {volume} {139}},\
  \bibinfo {pages} {084112} (\bibinfo {year} {2013})}\BibitemShut {NoStop}%
\bibitem [{\citenamefont {Mihm}, \citenamefont {Yang},\ and\ \citenamefont
  {Shepherd}(2020)}]{mihm_advanced_2020}%
  \BibitemOpen
  \bibfield  {author} {\bibinfo {author} {\bibfnamefont {T.~N.}\ \bibnamefont
  {Mihm}}, \bibinfo {author} {\bibfnamefont {B.}~\bibnamefont {Yang}}, \ and\
  \bibinfo {author} {\bibfnamefont {J.~J.}\ \bibnamefont {Shepherd}},\ }\href
  {http://arxiv.org/abs/2007.11696} {\bibfield  {journal} {\bibinfo  {journal}
  {arXiv:2007.11696 [cond-mat, physics:physics]}\ } (\bibinfo {year} {2020})},\
  \bibinfo {note} {arXiv: 2007.11696}\BibitemShut {NoStop}%
\bibitem [{\citenamefont {Shepherd}(2016)}]{shepherd_communication:_2016}%
  \BibitemOpen
  \bibfield  {author} {\bibinfo {author} {\bibfnamefont {J.~J.}\ \bibnamefont
  {Shepherd}},\ }\href {\doibase 10.1063/1.4958461} {\bibfield  {journal}
  {\bibinfo  {journal} {The Journal of Chemical Physics}\ }\textbf {\bibinfo
  {volume} {145}},\ \bibinfo {pages} {031104} (\bibinfo {year}
  {2016})}\BibitemShut {NoStop}%
\bibitem [{\citenamefont {Shepherd}\ \emph
  {et~al.}(2012{\natexlab{a}})\citenamefont {Shepherd}, \citenamefont
  {Grüneis}, \citenamefont {Booth}, \citenamefont {Kresse},\ and\
  \citenamefont {Alavi}}]{shepherd_convergence_2012}%
  \BibitemOpen
  \bibfield  {author} {\bibinfo {author} {\bibfnamefont {J.~J.}\ \bibnamefont
  {Shepherd}}, \bibinfo {author} {\bibfnamefont {A.}~\bibnamefont {Grüneis}},
  \bibinfo {author} {\bibfnamefont {G.~H.}\ \bibnamefont {Booth}}, \bibinfo
  {author} {\bibfnamefont {G.}~\bibnamefont {Kresse}}, \ and\ \bibinfo {author}
  {\bibfnamefont {A.}~\bibnamefont {Alavi}},\ }\href {\doibase
  10.1103/PhysRevB.86.035111} {\bibfield  {journal} {\bibinfo  {journal}
  {Physical Review B}\ }\textbf {\bibinfo {volume} {86}},\ \bibinfo {pages}
  {035111} (\bibinfo {year} {2012}{\natexlab{a}})}\BibitemShut {NoStop}%
\bibitem [{\citenamefont {Shepherd}, \citenamefont {Henderson},\ and\
  \citenamefont {Scuseria}(2014{\natexlab{a}})}]{shepherd_coupled_2014}%
  \BibitemOpen
  \bibfield  {author} {\bibinfo {author} {\bibfnamefont {J.~J.}\ \bibnamefont
  {Shepherd}}, \bibinfo {author} {\bibfnamefont {T.~M.}\ \bibnamefont
  {Henderson}}, \ and\ \bibinfo {author} {\bibfnamefont {G.~E.}\ \bibnamefont
  {Scuseria}},\ }\href {\doibase 10.1063/1.4867783} {\bibfield  {journal}
  {\bibinfo  {journal} {The Journal of Chemical Physics}\ }\textbf {\bibinfo
  {volume} {140}},\ \bibinfo {pages} {124102} (\bibinfo {year}
  {2014}{\natexlab{a}})}\BibitemShut {NoStop}%
\bibitem [{\citenamefont {Holzmann}, \citenamefont {Bernu},\ and\ \citenamefont
  {Ceperley}(2011)}]{holzmann_finite-size_2011}%
  \BibitemOpen
  \bibfield  {author} {\bibinfo {author} {\bibfnamefont {M.}~\bibnamefont
  {Holzmann}}, \bibinfo {author} {\bibfnamefont {B.}~\bibnamefont {Bernu}}, \
  and\ \bibinfo {author} {\bibfnamefont {D.~M.}\ \bibnamefont {Ceperley}},\
  }\href {\doibase 10.1088/1742-6596/321/1/012020} {\bibfield  {journal}
  {\bibinfo  {journal} {Journal of Physics: Conference Series}\ }\textbf
  {\bibinfo {volume} {321}},\ \bibinfo {pages} {012020} (\bibinfo {year}
  {2011})},\ \bibinfo {note} {publisher: IOP Publishing}\BibitemShut {NoStop}%
\bibitem [{\citenamefont {Sun}\ \emph {et~al.}(2018{\natexlab{a}})\citenamefont
  {Sun}, \citenamefont {Berkelbach}, \citenamefont {Blunt}, \citenamefont
  {Booth}, \citenamefont {Guo}, \citenamefont {Li}, \citenamefont {Liu},
  \citenamefont {McClain}, \citenamefont {Sayfutyarova}, \citenamefont
  {Sharma}, \citenamefont {Wouters},\ and\ \citenamefont
  {Chan}}]{sun_pyscf:_2018}%
  \BibitemOpen
  \bibfield  {author} {\bibinfo {author} {\bibfnamefont {Q.}~\bibnamefont
  {Sun}}, \bibinfo {author} {\bibfnamefont {T.~C.}\ \bibnamefont {Berkelbach}},
  \bibinfo {author} {\bibfnamefont {N.~S.}\ \bibnamefont {Blunt}}, \bibinfo
  {author} {\bibfnamefont {G.~H.}\ \bibnamefont {Booth}}, \bibinfo {author}
  {\bibfnamefont {S.}~\bibnamefont {Guo}}, \bibinfo {author} {\bibfnamefont
  {Z.}~\bibnamefont {Li}}, \bibinfo {author} {\bibfnamefont {J.}~\bibnamefont
  {Liu}}, \bibinfo {author} {\bibfnamefont {J.~D.}\ \bibnamefont {McClain}},
  \bibinfo {author} {\bibfnamefont {E.~R.}\ \bibnamefont {Sayfutyarova}},
  \bibinfo {author} {\bibfnamefont {S.}~\bibnamefont {Sharma}}, \bibinfo
  {author} {\bibfnamefont {S.}~\bibnamefont {Wouters}}, \ and\ \bibinfo
  {author} {\bibfnamefont {G.~K.-L.}\ \bibnamefont {Chan}},\ }\href {\doibase
  10.1002/wcms.1340} {\bibfield  {journal} {\bibinfo  {journal} {Wiley
  Interdisciplinary Reviews: Computational Molecular Science}\ }\textbf
  {\bibinfo {volume} {8}},\ \bibinfo {pages} {e1340} (\bibinfo {year}
  {2018}{\natexlab{a}})}\BibitemShut {NoStop}%
\bibitem [{\citenamefont {Sun}\ \emph {et~al.}(2018{\natexlab{b}})\citenamefont
  {Sun}, \citenamefont {Berkelbach}, \citenamefont {Blunt}, \citenamefont
  {Booth}, \citenamefont {Guo}, \citenamefont {Li}, \citenamefont {Liu},
  \citenamefont {McClain}, \citenamefont {Sayfutyarova}, \citenamefont
  {Sharma}, \citenamefont {Wouters},\ and\ \citenamefont
  {Chan}}]{sun_pyscf_2018}%
  \BibitemOpen
  \bibfield  {author} {\bibinfo {author} {\bibfnamefont {Q.}~\bibnamefont
  {Sun}}, \bibinfo {author} {\bibfnamefont {T.~C.}\ \bibnamefont {Berkelbach}},
  \bibinfo {author} {\bibfnamefont {N.~S.}\ \bibnamefont {Blunt}}, \bibinfo
  {author} {\bibfnamefont {G.~H.}\ \bibnamefont {Booth}}, \bibinfo {author}
  {\bibfnamefont {S.}~\bibnamefont {Guo}}, \bibinfo {author} {\bibfnamefont
  {Z.}~\bibnamefont {Li}}, \bibinfo {author} {\bibfnamefont {J.}~\bibnamefont
  {Liu}}, \bibinfo {author} {\bibfnamefont {J.~D.}\ \bibnamefont {McClain}},
  \bibinfo {author} {\bibfnamefont {E.~R.}\ \bibnamefont {Sayfutyarova}},
  \bibinfo {author} {\bibfnamefont {S.}~\bibnamefont {Sharma}}, \bibinfo
  {author} {\bibfnamefont {S.}~\bibnamefont {Wouters}}, \ and\ \bibinfo
  {author} {\bibfnamefont {G.~K.}\ \bibnamefont {Chan}},\ }\href {\doibase
  10.1002/wcms.1340} {\bibfield  {journal} {\bibinfo  {journal} {WIREs
  Computational Molecular Science}\ }\textbf {\bibinfo {volume} {8}} (\bibinfo
  {year} {2018}{\natexlab{b}}),\ 10.1002/wcms.1340}\BibitemShut {NoStop}%
\bibitem [{\citenamefont {Wang}\ and\ \citenamefont
  {Berkelbach}(2020)}]{wang_excitons_2020}%
  \BibitemOpen
  \bibfield  {author} {\bibinfo {author} {\bibfnamefont {X.}~\bibnamefont
  {Wang}}\ and\ \bibinfo {author} {\bibfnamefont {T.~C.}\ \bibnamefont
  {Berkelbach}},\ }\href {\doibase 10.1021/acs.jctc.0c00101} {\bibfield
  {journal} {\bibinfo  {journal} {Journal of Chemical Theory and Computation}\
  }\textbf {\bibinfo {volume} {16}},\ \bibinfo {pages} {3095} (\bibinfo {year}
  {2020})},\ \bibinfo {note} {publisher: American Chemical Society}\BibitemShut
  {NoStop}%
\bibitem [{\citenamefont {McClain}\ \emph {et~al.}(2017)\citenamefont
  {McClain}, \citenamefont {Sun}, \citenamefont {Chan},\ and\ \citenamefont
  {Berkelbach}}]{mcclain_gaussian-based_2017}%
  \BibitemOpen
  \bibfield  {author} {\bibinfo {author} {\bibfnamefont {J.}~\bibnamefont
  {McClain}}, \bibinfo {author} {\bibfnamefont {Q.}~\bibnamefont {Sun}},
  \bibinfo {author} {\bibfnamefont {G.~K.-L.}\ \bibnamefont {Chan}}, \ and\
  \bibinfo {author} {\bibfnamefont {T.~C.}\ \bibnamefont {Berkelbach}},\ }\href
  {\doibase 10.1021/acs.jctc.7b00049} {\bibfield  {journal} {\bibinfo
  {journal} {Journal of Chemical Theory and Computation}\ }\textbf {\bibinfo
  {volume} {13}},\ \bibinfo {pages} {1209} (\bibinfo {year}
  {2017})}\BibitemShut {NoStop}%
\bibitem [{\citenamefont {McClain}\ \emph {et~al.}(2016)\citenamefont
  {McClain}, \citenamefont {Lischner}, \citenamefont {Watson}, \citenamefont
  {Matthews}, \citenamefont {Ronca}, \citenamefont {Louie}, \citenamefont
  {Berkelbach},\ and\ \citenamefont {Chan}}]{mcclain_spectral_2016}%
  \BibitemOpen
  \bibfield  {author} {\bibinfo {author} {\bibfnamefont {J.}~\bibnamefont
  {McClain}}, \bibinfo {author} {\bibfnamefont {J.}~\bibnamefont {Lischner}},
  \bibinfo {author} {\bibfnamefont {T.}~\bibnamefont {Watson}}, \bibinfo
  {author} {\bibfnamefont {D.~A.}\ \bibnamefont {Matthews}}, \bibinfo {author}
  {\bibfnamefont {E.}~\bibnamefont {Ronca}}, \bibinfo {author} {\bibfnamefont
  {S.~G.}\ \bibnamefont {Louie}}, \bibinfo {author} {\bibfnamefont {T.~C.}\
  \bibnamefont {Berkelbach}}, \ and\ \bibinfo {author} {\bibfnamefont
  {G.~K.-L.}\ \bibnamefont {Chan}},\ }\href {\doibase
  10.1103/PhysRevB.93.235139} {\bibfield  {journal} {\bibinfo  {journal}
  {Physical Review B}\ }\textbf {\bibinfo {volume} {93}},\ \bibinfo {pages}
  {235139} (\bibinfo {year} {2016})}\BibitemShut {NoStop}%
\bibitem [{\citenamefont {Lewis}\ and\ \citenamefont
  {Berkelbach}(2019)}]{lewis_ab_2019}%
  \BibitemOpen
  \bibfield  {author} {\bibinfo {author} {\bibfnamefont {A.~M.}\ \bibnamefont
  {Lewis}}\ and\ \bibinfo {author} {\bibfnamefont {T.~C.}\ \bibnamefont
  {Berkelbach}},\ }\href {\doibase 10.1103/PhysRevLett.122.226402} {\bibfield
  {journal} {\bibinfo  {journal} {Physical Review Letters}\ }\textbf {\bibinfo
  {volume} {122}},\ \bibinfo {pages} {226402} (\bibinfo {year}
  {2019})}\BibitemShut {NoStop}%
\bibitem [{\citenamefont {Booth}\ \emph {et~al.}(2016)\citenamefont {Booth},
  \citenamefont {Tsatsoulis}, \citenamefont {Chan},\ and\ \citenamefont
  {Grüneis}}]{booth_plane_2016}%
  \BibitemOpen
  \bibfield  {author} {\bibinfo {author} {\bibfnamefont {G.~H.}\ \bibnamefont
  {Booth}}, \bibinfo {author} {\bibfnamefont {T.}~\bibnamefont {Tsatsoulis}},
  \bibinfo {author} {\bibfnamefont {G.~K.-L.}\ \bibnamefont {Chan}}, \ and\
  \bibinfo {author} {\bibfnamefont {A.}~\bibnamefont {Grüneis}},\ }\href
  {http://arxiv.org/abs/1603.06457} {\bibfield  {journal} {\bibinfo  {journal}
  {arXiv:1603.06457 [cond-mat, physics:physics]}\ } (\bibinfo {year} {2016})},\
  \bibinfo {note} {arXiv: 1603.06457}\BibitemShut {NoStop}%
\bibitem [{\citenamefont {Dornheim}\ \emph {et~al.}(2016)\citenamefont
  {Dornheim}, \citenamefont {Groth}, \citenamefont {Sjostrom}, \citenamefont
  {Malone}, \citenamefont {Foulkes},\ and\ \citenamefont
  {Bonitz}}]{dornheim_ab_2016}%
  \BibitemOpen
  \bibfield  {author} {\bibinfo {author} {\bibfnamefont {T.}~\bibnamefont
  {Dornheim}}, \bibinfo {author} {\bibfnamefont {S.}~\bibnamefont {Groth}},
  \bibinfo {author} {\bibfnamefont {T.}~\bibnamefont {Sjostrom}}, \bibinfo
  {author} {\bibfnamefont {F.~D.}\ \bibnamefont {Malone}}, \bibinfo {author}
  {\bibfnamefont {W.}~\bibnamefont {Foulkes}}, \ and\ \bibinfo {author}
  {\bibfnamefont {M.}~\bibnamefont {Bonitz}},\ }\href {\doibase
  10.1103/PhysRevLett.117.156403} {\bibfield  {journal} {\bibinfo  {journal}
  {Physical Review Letters}\ }\textbf {\bibinfo {volume} {117}},\ \bibinfo
  {pages} {156403} (\bibinfo {year} {2016})}\BibitemShut {NoStop}%
\bibitem [{\citenamefont {Harl}, \citenamefont {Schimka},\ and\ \citenamefont
  {Kresse}(2010)}]{harl_assessing_2010}%
  \BibitemOpen
  \bibfield  {author} {\bibinfo {author} {\bibfnamefont {J.}~\bibnamefont
  {Harl}}, \bibinfo {author} {\bibfnamefont {L.}~\bibnamefont {Schimka}}, \
  and\ \bibinfo {author} {\bibfnamefont {G.}~\bibnamefont {Kresse}},\ }\href
  {\doibase 10.1103/PhysRevB.81.115126} {\bibfield  {journal} {\bibinfo
  {journal} {Physical Review B}\ }\textbf {\bibinfo {volume} {81}},\ \bibinfo
  {pages} {115126} (\bibinfo {year} {2010})}\BibitemShut {NoStop}%
\bibitem [{\citenamefont {Lebègue}\ \emph {et~al.}(2010)\citenamefont
  {Lebègue}, \citenamefont {Harl}, \citenamefont {Gould}, \citenamefont
  {Ángyán}, \citenamefont {Kresse},\ and\ \citenamefont
  {Dobson}}]{lebegue_cohesive_2010}%
  \BibitemOpen
  \bibfield  {author} {\bibinfo {author} {\bibfnamefont {S.}~\bibnamefont
  {Lebègue}}, \bibinfo {author} {\bibfnamefont {J.}~\bibnamefont {Harl}},
  \bibinfo {author} {\bibfnamefont {T.}~\bibnamefont {Gould}}, \bibinfo
  {author} {\bibfnamefont {J.~G.}\ \bibnamefont {Ángyán}}, \bibinfo {author}
  {\bibfnamefont {G.}~\bibnamefont {Kresse}}, \ and\ \bibinfo {author}
  {\bibfnamefont {J.~F.}\ \bibnamefont {Dobson}},\ }\href {\doibase
  10.1103/PhysRevLett.105.196401} {\bibfield  {journal} {\bibinfo  {journal}
  {Physical Review Letters}\ }\textbf {\bibinfo {volume} {105}},\ \bibinfo
  {pages} {196401} (\bibinfo {year} {2010})}\BibitemShut {NoStop}%
\bibitem [{\citenamefont {Schimka}\ \emph {et~al.}(2010)\citenamefont
  {Schimka}, \citenamefont {Harl}, \citenamefont {Stroppa}, \citenamefont
  {Grüneis}, \citenamefont {Marsman}, \citenamefont {Mittendorfer},\ and\
  \citenamefont {Kresse}}]{schimka_accurate_2010}%
  \BibitemOpen
  \bibfield  {author} {\bibinfo {author} {\bibfnamefont {L.}~\bibnamefont
  {Schimka}}, \bibinfo {author} {\bibfnamefont {J.}~\bibnamefont {Harl}},
  \bibinfo {author} {\bibfnamefont {A.}~\bibnamefont {Stroppa}}, \bibinfo
  {author} {\bibfnamefont {A.}~\bibnamefont {Grüneis}}, \bibinfo {author}
  {\bibfnamefont {M.}~\bibnamefont {Marsman}}, \bibinfo {author} {\bibfnamefont
  {F.}~\bibnamefont {Mittendorfer}}, \ and\ \bibinfo {author} {\bibfnamefont
  {G.}~\bibnamefont {Kresse}},\ }\href {\doibase 10.1038/nmat2806} {\bibfield
  {journal} {\bibinfo  {journal} {Nature Materials}\ }\textbf {\bibinfo
  {volume} {9}},\ \bibinfo {pages} {741} (\bibinfo {year} {2010})},\ \bibinfo
  {note} {number: 9 Publisher: Nature Publishing Group}\BibitemShut {NoStop}%
\bibitem [{\citenamefont {Harl}\ and\ \citenamefont
  {Kresse}(2009)}]{harl_accurate_2009}%
  \BibitemOpen
  \bibfield  {author} {\bibinfo {author} {\bibfnamefont {J.}~\bibnamefont
  {Harl}}\ and\ \bibinfo {author} {\bibfnamefont {G.}~\bibnamefont {Kresse}},\
  }\href {\doibase 10.1103/PhysRevLett.103.056401} {\bibfield  {journal}
  {\bibinfo  {journal} {Physical Review Letters}\ }\textbf {\bibinfo {volume}
  {103}},\ \bibinfo {pages} {056401} (\bibinfo {year} {2009})}\BibitemShut
  {NoStop}%
\bibitem [{\citenamefont {Grüneis}\ \emph {et~al.}(2009)\citenamefont
  {Grüneis}, \citenamefont {Marsman}, \citenamefont {Harl}, \citenamefont
  {Schimka},\ and\ \citenamefont {Kresse}}]{gruneis_making_2009}%
  \BibitemOpen
  \bibfield  {author} {\bibinfo {author} {\bibfnamefont {A.}~\bibnamefont
  {Grüneis}}, \bibinfo {author} {\bibfnamefont {M.}~\bibnamefont {Marsman}},
  \bibinfo {author} {\bibfnamefont {J.}~\bibnamefont {Harl}}, \bibinfo {author}
  {\bibfnamefont {L.}~\bibnamefont {Schimka}}, \ and\ \bibinfo {author}
  {\bibfnamefont {G.}~\bibnamefont {Kresse}},\ }\href {\doibase
  10.1063/1.3250347} {\bibfield  {journal} {\bibinfo  {journal} {The Journal of
  Chemical Physics}\ }\textbf {\bibinfo {volume} {131}},\ \bibinfo {pages}
  {154115} (\bibinfo {year} {2009})}\BibitemShut {NoStop}%
\bibitem [{\citenamefont {Riemelmoser}, \citenamefont {Kaltak},\ and\
  \citenamefont {Kresse}(2020)}]{riemelmoser_plane_2020}%
  \BibitemOpen
  \bibfield  {author} {\bibinfo {author} {\bibfnamefont {S.}~\bibnamefont
  {Riemelmoser}}, \bibinfo {author} {\bibfnamefont {M.}~\bibnamefont {Kaltak}},
  \ and\ \bibinfo {author} {\bibfnamefont {G.}~\bibnamefont {Kresse}},\ }\href
  {\doibase 10.1063/5.0002246} {\bibfield  {journal} {\bibinfo  {journal} {The
  Journal of Chemical Physics}\ }\textbf {\bibinfo {volume} {152}},\ \bibinfo
  {pages} {134103} (\bibinfo {year} {2020})},\ \bibinfo {note} {publisher:
  American Institute of Physics}\BibitemShut {NoStop}%
\bibitem [{\citenamefont {Motta}\ \emph {et~al.}(2017)\citenamefont {Motta},
  \citenamefont {Ceperley}, \citenamefont {Chan}, \citenamefont {Gomez},
  \citenamefont {Gull}, \citenamefont {Guo}, \citenamefont {Jiménez-Hoyos},
  \citenamefont {Lan}, \citenamefont {Li}, \citenamefont {Ma}, \citenamefont
  {Millis}, \citenamefont {Prokof’ev}, \citenamefont {Ray}, \citenamefont
  {Scuseria}, \citenamefont {Sorella}, \citenamefont {Stoudenmire},
  \citenamefont {Sun}, \citenamefont {Tupitsyn}, \citenamefont {White},
  \citenamefont {Zgid}, \citenamefont {Zhang},\ and\ \citenamefont {{Simons
  Collaboration on the Many-Electron Problem}}}]{motta_towards_2017}%
  \BibitemOpen
  \bibfield  {author} {\bibinfo {author} {\bibfnamefont {M.}~\bibnamefont
  {Motta}}, \bibinfo {author} {\bibfnamefont {D.~M.}\ \bibnamefont {Ceperley}},
  \bibinfo {author} {\bibfnamefont {G.~K.-L.}\ \bibnamefont {Chan}}, \bibinfo
  {author} {\bibfnamefont {J.~A.}\ \bibnamefont {Gomez}}, \bibinfo {author}
  {\bibfnamefont {E.}~\bibnamefont {Gull}}, \bibinfo {author} {\bibfnamefont
  {S.}~\bibnamefont {Guo}}, \bibinfo {author} {\bibfnamefont {C.~A.}\
  \bibnamefont {Jiménez-Hoyos}}, \bibinfo {author} {\bibfnamefont {T.~N.}\
  \bibnamefont {Lan}}, \bibinfo {author} {\bibfnamefont {J.}~\bibnamefont
  {Li}}, \bibinfo {author} {\bibfnamefont {F.}~\bibnamefont {Ma}}, \bibinfo
  {author} {\bibfnamefont {A.~J.}\ \bibnamefont {Millis}}, \bibinfo {author}
  {\bibfnamefont {N.~V.}\ \bibnamefont {Prokof’ev}}, \bibinfo {author}
  {\bibfnamefont {U.}~\bibnamefont {Ray}}, \bibinfo {author} {\bibfnamefont
  {G.~E.}\ \bibnamefont {Scuseria}}, \bibinfo {author} {\bibfnamefont
  {S.}~\bibnamefont {Sorella}}, \bibinfo {author} {\bibfnamefont {E.~M.}\
  \bibnamefont {Stoudenmire}}, \bibinfo {author} {\bibfnamefont
  {Q.}~\bibnamefont {Sun}}, \bibinfo {author} {\bibfnamefont {I.~S.}\
  \bibnamefont {Tupitsyn}}, \bibinfo {author} {\bibfnamefont {S.~R.}\
  \bibnamefont {White}}, \bibinfo {author} {\bibfnamefont {D.}~\bibnamefont
  {Zgid}}, \bibinfo {author} {\bibfnamefont {S.}~\bibnamefont {Zhang}}, \ and\
  \bibinfo {author} {\bibnamefont {{Simons Collaboration on the Many-Electron
  Problem}}},\ }\href {\doibase 10.1103/PhysRevX.7.031059} {\bibfield
  {journal} {\bibinfo  {journal} {Physical Review X}\ }\textbf {\bibinfo
  {volume} {7}},\ \bibinfo {pages} {031059} (\bibinfo {year}
  {2017})}\BibitemShut {NoStop}%
\bibitem [{\citenamefont {Grüneis}(2015)}]{gruneis_efficient_2015}%
  \BibitemOpen
  \bibfield  {author} {\bibinfo {author} {\bibfnamefont {A.}~\bibnamefont
  {Grüneis}},\ }\href {\doibase 10.1103/PhysRevLett.115.066402} {\bibfield
  {journal} {\bibinfo  {journal} {Physical Review Letters}\ }\textbf {\bibinfo
  {volume} {115}},\ \bibinfo {pages} {066402} (\bibinfo {year}
  {2015})}\BibitemShut {NoStop}%
\bibitem [{\citenamefont {Grüneis}, \citenamefont {Marsman},\ and\
  \citenamefont {Kresse}(2010)}]{gruneis_second-order_2010}%
  \BibitemOpen
  \bibfield  {author} {\bibinfo {author} {\bibfnamefont {A.}~\bibnamefont
  {Grüneis}}, \bibinfo {author} {\bibfnamefont {M.}~\bibnamefont {Marsman}}, \
  and\ \bibinfo {author} {\bibfnamefont {G.}~\bibnamefont {Kresse}},\ }\href
  {\doibase 10.1063/1.3466765} {\bibfield  {journal} {\bibinfo  {journal} {The
  Journal of Chemical Physics}\ }\textbf {\bibinfo {volume} {133}},\ \bibinfo
  {pages} {074107} (\bibinfo {year} {2010})}\BibitemShut {NoStop}%
\bibitem [{\citenamefont {Irmler}\ \emph {et~al.}(2019)\citenamefont {Irmler},
  \citenamefont {Gallo}, \citenamefont {Hummel},\ and\ \citenamefont
  {Grüneis}}]{irmler_duality_2019}%
  \BibitemOpen
  \bibfield  {author} {\bibinfo {author} {\bibfnamefont {A.}~\bibnamefont
  {Irmler}}, \bibinfo {author} {\bibfnamefont {A.}~\bibnamefont {Gallo}},
  \bibinfo {author} {\bibfnamefont {F.}~\bibnamefont {Hummel}}, \ and\ \bibinfo
  {author} {\bibfnamefont {A.}~\bibnamefont {Grüneis}},\ }\href {\doibase
  10.1103/PhysRevLett.123.156401} {\bibfield  {journal} {\bibinfo  {journal}
  {Physical Review Letters}\ }\textbf {\bibinfo {volume} {123}},\ \bibinfo
  {pages} {156401} (\bibinfo {year} {2019})}\BibitemShut {NoStop}%
\bibitem [{\citenamefont {Drummond}\ \emph {et~al.}(2008)\citenamefont
  {Drummond}, \citenamefont {Needs}, \citenamefont {Sorouri},\ and\
  \citenamefont {Foulkes}}]{drummond_finite-size_2008}%
  \BibitemOpen
  \bibfield  {author} {\bibinfo {author} {\bibfnamefont {N.~D.}\ \bibnamefont
  {Drummond}}, \bibinfo {author} {\bibfnamefont {R.~J.}\ \bibnamefont {Needs}},
  \bibinfo {author} {\bibfnamefont {A.}~\bibnamefont {Sorouri}}, \ and\
  \bibinfo {author} {\bibfnamefont {W.~M.~C.}\ \bibnamefont {Foulkes}},\ }\href
  {\doibase 10.1103/PhysRevB.78.125106} {\bibfield  {journal} {\bibinfo
  {journal} {Physical Review B}\ }\textbf {\bibinfo {volume} {78}},\ \bibinfo
  {pages} {125106} (\bibinfo {year} {2008})}\BibitemShut {NoStop}%
\bibitem [{\citenamefont {Lin}, \citenamefont {Zong},\ and\ \citenamefont
  {Ceperley}(2001)}]{lin_twist-averaged_2001}%
  \BibitemOpen
  \bibfield  {author} {\bibinfo {author} {\bibfnamefont {C.}~\bibnamefont
  {Lin}}, \bibinfo {author} {\bibfnamefont {F.~H.}\ \bibnamefont {Zong}}, \
  and\ \bibinfo {author} {\bibfnamefont {D.~M.}\ \bibnamefont {Ceperley}},\
  }\href {\doibase 10.1103/PhysRevE.64.016702} {\bibfield  {journal} {\bibinfo
  {journal} {Physical Review E}\ }\textbf {\bibinfo {volume} {64}},\ \bibinfo
  {pages} {016702} (\bibinfo {year} {2001})}\BibitemShut {NoStop}%
\bibitem [{\citenamefont {Chiesa}\ \emph {et~al.}(2006)\citenamefont {Chiesa},
  \citenamefont {Ceperley}, \citenamefont {Martin},\ and\ \citenamefont
  {Holzmann}}]{chiesa_finite-size_2006}%
  \BibitemOpen
  \bibfield  {author} {\bibinfo {author} {\bibfnamefont {S.}~\bibnamefont
  {Chiesa}}, \bibinfo {author} {\bibfnamefont {D.~M.}\ \bibnamefont
  {Ceperley}}, \bibinfo {author} {\bibfnamefont {R.~M.}\ \bibnamefont
  {Martin}}, \ and\ \bibinfo {author} {\bibfnamefont {M.}~\bibnamefont
  {Holzmann}},\ }\href {\doibase 10.1103/PhysRevLett.97.076404} {\bibfield
  {journal} {\bibinfo  {journal} {Physical Review Letters}\ }\textbf {\bibinfo
  {volume} {97}},\ \bibinfo {pages} {076404} (\bibinfo {year}
  {2006})}\BibitemShut {NoStop}%
\bibitem [{\citenamefont {Mostaani}, \citenamefont {Drummond},\ and\
  \citenamefont {Fal’ko}(2015)}]{mostaani_quantum_2015}%
  \BibitemOpen
  \bibfield  {author} {\bibinfo {author} {\bibfnamefont {E.}~\bibnamefont
  {Mostaani}}, \bibinfo {author} {\bibfnamefont {N.}~\bibnamefont {Drummond}},
  \ and\ \bibinfo {author} {\bibfnamefont {V.}~\bibnamefont {Fal’ko}},\
  }\href {\doibase 10.1103/PhysRevLett.115.115501} {\bibfield  {journal}
  {\bibinfo  {journal} {Physical Review Letters}\ }\textbf {\bibinfo {volume}
  {115}},\ \bibinfo {pages} {115501} (\bibinfo {year} {2015})}\BibitemShut
  {NoStop}%
\bibitem [{\citenamefont {Pierleoni}, \citenamefont {Ceperley},\ and\
  \citenamefont {Holzmann}(2004)}]{pierleoni_coupled_2004}%
  \BibitemOpen
  \bibfield  {author} {\bibinfo {author} {\bibfnamefont {C.}~\bibnamefont
  {Pierleoni}}, \bibinfo {author} {\bibfnamefont {D.~M.}\ \bibnamefont
  {Ceperley}}, \ and\ \bibinfo {author} {\bibfnamefont {M.}~\bibnamefont
  {Holzmann}},\ }\href {\doibase 10.1103/PhysRevLett.93.146402} {\bibfield
  {journal} {\bibinfo  {journal} {Physical Review Letters}\ }\textbf {\bibinfo
  {volume} {93}},\ \bibinfo {pages} {146402} (\bibinfo {year}
  {2004})}\BibitemShut {NoStop}%
\bibitem [{\citenamefont {Fraser}\ \emph {et~al.}(1996)\citenamefont {Fraser},
  \citenamefont {Foulkes}, \citenamefont {Rajagopal}, \citenamefont {Needs},
  \citenamefont {Kenny},\ and\ \citenamefont
  {Williamson}}]{fraser_finite-size_1996}%
  \BibitemOpen
  \bibfield  {author} {\bibinfo {author} {\bibfnamefont {L.~M.}\ \bibnamefont
  {Fraser}}, \bibinfo {author} {\bibfnamefont {W.~M.~C.}\ \bibnamefont
  {Foulkes}}, \bibinfo {author} {\bibfnamefont {G.}~\bibnamefont {Rajagopal}},
  \bibinfo {author} {\bibfnamefont {R.~J.}\ \bibnamefont {Needs}}, \bibinfo
  {author} {\bibfnamefont {S.~D.}\ \bibnamefont {Kenny}}, \ and\ \bibinfo
  {author} {\bibfnamefont {A.~J.}\ \bibnamefont {Williamson}},\ }\href
  {\doibase 10.1103/PhysRevB.53.1814} {\bibfield  {journal} {\bibinfo
  {journal} {Physical Review B}\ }\textbf {\bibinfo {volume} {53}},\ \bibinfo
  {pages} {1814} (\bibinfo {year} {1996})}\BibitemShut {NoStop}%
\bibitem [{\citenamefont {Williamson}\ \emph {et~al.}(1997)\citenamefont
  {Williamson}, \citenamefont {Rajagopal}, \citenamefont {Needs}, \citenamefont
  {Fraser}, \citenamefont {Foulkes}, \citenamefont {Wang},\ and\ \citenamefont
  {Chou}}]{williamson_elimination_1997}%
  \BibitemOpen
  \bibfield  {author} {\bibinfo {author} {\bibfnamefont {A.~J.}\ \bibnamefont
  {Williamson}}, \bibinfo {author} {\bibfnamefont {G.}~\bibnamefont
  {Rajagopal}}, \bibinfo {author} {\bibfnamefont {R.~J.}\ \bibnamefont
  {Needs}}, \bibinfo {author} {\bibfnamefont {L.~M.}\ \bibnamefont {Fraser}},
  \bibinfo {author} {\bibfnamefont {W.~M.~C.}\ \bibnamefont {Foulkes}},
  \bibinfo {author} {\bibfnamefont {Y.}~\bibnamefont {Wang}}, \ and\ \bibinfo
  {author} {\bibfnamefont {M.-Y.}\ \bibnamefont {Chou}},\ }\href {\doibase
  10.1103/PhysRevB.55.R4851} {\bibfield  {journal} {\bibinfo  {journal}
  {Physical Review B}\ }\textbf {\bibinfo {volume} {55}},\ \bibinfo {pages}
  {R4851} (\bibinfo {year} {1997})}\BibitemShut {NoStop}%
\bibitem [{\citenamefont {Zong}, \citenamefont {Lin},\ and\ \citenamefont
  {Ceperley}(2002)}]{zong_spin_2002}%
  \BibitemOpen
  \bibfield  {author} {\bibinfo {author} {\bibfnamefont {F.~H.}\ \bibnamefont
  {Zong}}, \bibinfo {author} {\bibfnamefont {C.}~\bibnamefont {Lin}}, \ and\
  \bibinfo {author} {\bibfnamefont {D.~M.}\ \bibnamefont {Ceperley}},\ }\href
  {\doibase 10.1103/PhysRevE.66.036703} {\bibfield  {journal} {\bibinfo
  {journal} {Physical Review E}\ }\textbf {\bibinfo {volume} {66}},\ \bibinfo
  {pages} {036703} (\bibinfo {year} {2002})}\BibitemShut {NoStop}%
\bibitem [{\citenamefont {Filippi}\ and\ \citenamefont
  {Ceperley}(1999)}]{filippi_quantum_1999}%
  \BibitemOpen
  \bibfield  {author} {\bibinfo {author} {\bibfnamefont {C.}~\bibnamefont
  {Filippi}}\ and\ \bibinfo {author} {\bibfnamefont {D.~M.}\ \bibnamefont
  {Ceperley}},\ }\href {\doibase 10.1103/PhysRevB.59.7907} {\bibfield
  {journal} {\bibinfo  {journal} {Physical Review B}\ }\textbf {\bibinfo
  {volume} {59}},\ \bibinfo {pages} {7907} (\bibinfo {year} {1999})},\ \bibinfo
  {note} {publisher: American Physical Society}\BibitemShut {NoStop}%
\bibitem [{\citenamefont {Baldereschi}(1973)}]{baldereschi_mean-value_1973}%
  \BibitemOpen
  \bibfield  {author} {\bibinfo {author} {\bibfnamefont {A.}~\bibnamefont
  {Baldereschi}},\ }\href {\doibase 10.1103/PhysRevB.7.5212} {\bibfield
  {journal} {\bibinfo  {journal} {Physical Review B}\ }\textbf {\bibinfo
  {volume} {7}},\ \bibinfo {pages} {5212} (\bibinfo {year} {1973})}\BibitemShut
  {NoStop}%
\bibitem [{\citenamefont {Rajagopal}\ \emph {et~al.}(1995)\citenamefont
  {Rajagopal}, \citenamefont {Needs}, \citenamefont {James}, \citenamefont
  {Kenny},\ and\ \citenamefont {Foulkes}}]{rajagopal_variational_1995}%
  \BibitemOpen
  \bibfield  {author} {\bibinfo {author} {\bibfnamefont {G.}~\bibnamefont
  {Rajagopal}}, \bibinfo {author} {\bibfnamefont {R.~J.}\ \bibnamefont
  {Needs}}, \bibinfo {author} {\bibfnamefont {A.}~\bibnamefont {James}},
  \bibinfo {author} {\bibfnamefont {S.~D.}\ \bibnamefont {Kenny}}, \ and\
  \bibinfo {author} {\bibfnamefont {W.~M.~C.}\ \bibnamefont {Foulkes}},\ }\href
  {\doibase 10.1103/PhysRevB.51.10591} {\bibfield  {journal} {\bibinfo
  {journal} {Physical Review B}\ }\textbf {\bibinfo {volume} {51}},\ \bibinfo
  {pages} {10591} (\bibinfo {year} {1995})}\BibitemShut {NoStop}%
\bibitem [{\citenamefont {Kent}\ \emph {et~al.}(1999)\citenamefont {Kent},
  \citenamefont {Hood}, \citenamefont {Williamson}, \citenamefont {Needs},
  \citenamefont {Foulkes},\ and\ \citenamefont
  {Rajagopal}}]{kent_finite-size_1999}%
  \BibitemOpen
  \bibfield  {author} {\bibinfo {author} {\bibfnamefont {P.~R.~C.}\
  \bibnamefont {Kent}}, \bibinfo {author} {\bibfnamefont {R.~Q.}\ \bibnamefont
  {Hood}}, \bibinfo {author} {\bibfnamefont {A.~J.}\ \bibnamefont
  {Williamson}}, \bibinfo {author} {\bibfnamefont {R.~J.}\ \bibnamefont
  {Needs}}, \bibinfo {author} {\bibfnamefont {W.~M.~C.}\ \bibnamefont
  {Foulkes}}, \ and\ \bibinfo {author} {\bibfnamefont {G.}~\bibnamefont
  {Rajagopal}},\ }\href {\doibase 10.1103/PhysRevB.59.1917} {\bibfield
  {journal} {\bibinfo  {journal} {Physical Review B}\ }\textbf {\bibinfo
  {volume} {59}},\ \bibinfo {pages} {1917} (\bibinfo {year}
  {1999})}\BibitemShut {NoStop}%
\bibitem [{\citenamefont {Cleland}, \citenamefont {Booth},\ and\ \citenamefont
  {Alavi}(2010)}]{cleland_communications:_2010}%
  \BibitemOpen
  \bibfield  {author} {\bibinfo {author} {\bibfnamefont {D.}~\bibnamefont
  {Cleland}}, \bibinfo {author} {\bibfnamefont {G.~H.}\ \bibnamefont {Booth}},
  \ and\ \bibinfo {author} {\bibfnamefont {A.}~\bibnamefont {Alavi}},\ }\href
  {\doibase 10.1063/1.3302277} {\bibfield  {journal} {\bibinfo  {journal} {The
  Journal of Chemical Physics}\ }\textbf {\bibinfo {volume} {132}},\ \bibinfo
  {pages} {041103} (\bibinfo {year} {2010})}\BibitemShut {NoStop}%
\bibitem [{\citenamefont {Blunt}(2018)}]{blunt_communication_2018}%
  \BibitemOpen
  \bibfield  {author} {\bibinfo {author} {\bibfnamefont {N.~S.}\ \bibnamefont
  {Blunt}},\ }\href {\doibase 10.1063/1.5037923} {\bibfield  {journal}
  {\bibinfo  {journal} {The Journal of Chemical Physics}\ }\textbf {\bibinfo
  {volume} {148}},\ \bibinfo {pages} {221101} (\bibinfo {year}
  {2018})}\BibitemShut {NoStop}%
\bibitem [{\citenamefont {Shepherd}, \citenamefont {Booth},\ and\ \citenamefont
  {Alavi}(2012)}]{shepherd_investigation_2012}%
  \BibitemOpen
  \bibfield  {author} {\bibinfo {author} {\bibfnamefont {J.~J.}\ \bibnamefont
  {Shepherd}}, \bibinfo {author} {\bibfnamefont {G.~H.}\ \bibnamefont {Booth}},
  \ and\ \bibinfo {author} {\bibfnamefont {A.}~\bibnamefont {Alavi}},\ }\href
  {\doibase 10.1063/1.4720076} {\bibfield  {journal} {\bibinfo  {journal} {The
  Journal of Chemical Physics}\ }\textbf {\bibinfo {volume} {136}},\ \bibinfo
  {pages} {244101} (\bibinfo {year} {2012})}\BibitemShut {NoStop}%
\bibitem [{\citenamefont {Shepherd}, \citenamefont {Scuseria},\ and\
  \citenamefont {Spencer}(2014)}]{shepherd_sign_2014}%
  \BibitemOpen
  \bibfield  {author} {\bibinfo {author} {\bibfnamefont {J.~J.}\ \bibnamefont
  {Shepherd}}, \bibinfo {author} {\bibfnamefont {G.~E.}\ \bibnamefont
  {Scuseria}}, \ and\ \bibinfo {author} {\bibfnamefont {J.~S.}\ \bibnamefont
  {Spencer}},\ }\href {\doibase 10.1103/PhysRevB.90.155130} {\bibfield
  {journal} {\bibinfo  {journal} {Physical Review B}\ }\textbf {\bibinfo
  {volume} {90}},\ \bibinfo {pages} {155130} (\bibinfo {year}
  {2014})}\BibitemShut {NoStop}%
\bibitem [{\citenamefont {Blunt}(2019)}]{blunt_hybrid_2019}%
  \BibitemOpen
  \bibfield  {author} {\bibinfo {author} {\bibfnamefont {N.~S.}\ \bibnamefont
  {Blunt}},\ }\href {\doibase 10.1063/1.5123146} {\bibfield  {journal}
  {\bibinfo  {journal} {The Journal of Chemical Physics}\ }\textbf {\bibinfo
  {volume} {151}},\ \bibinfo {pages} {174103} (\bibinfo {year}
  {2019})}\BibitemShut {NoStop}%
\bibitem [{\citenamefont {Ghanem}, \citenamefont {Lozovoi},\ and\ \citenamefont
  {Alavi}(2019)}]{ghanem_unbiasing_2019}%
  \BibitemOpen
  \bibfield  {author} {\bibinfo {author} {\bibfnamefont {K.}~\bibnamefont
  {Ghanem}}, \bibinfo {author} {\bibfnamefont {A.~Y.}\ \bibnamefont {Lozovoi}},
  \ and\ \bibinfo {author} {\bibfnamefont {A.}~\bibnamefont {Alavi}},\ }\href
  {\doibase 10.1063/1.5134006} {\bibfield  {journal} {\bibinfo  {journal} {The
  Journal of Chemical Physics}\ }\textbf {\bibinfo {volume} {151}},\ \bibinfo
  {pages} {224108} (\bibinfo {year} {2019})}\BibitemShut {NoStop}%
\bibitem [{\citenamefont {Schwarz}, \citenamefont {Booth},\ and\ \citenamefont
  {Alavi}(2015)}]{schwarz_insights_2015}%
  \BibitemOpen
  \bibfield  {author} {\bibinfo {author} {\bibfnamefont {L.~R.}\ \bibnamefont
  {Schwarz}}, \bibinfo {author} {\bibfnamefont {G.~H.}\ \bibnamefont {Booth}},
  \ and\ \bibinfo {author} {\bibfnamefont {A.}~\bibnamefont {Alavi}},\ }\href
  {\doibase 10.1103/PhysRevB.91.045139} {\bibfield  {journal} {\bibinfo
  {journal} {Physical Review B}\ }\textbf {\bibinfo {volume} {91}},\ \bibinfo
  {pages} {045139} (\bibinfo {year} {2015})}\BibitemShut {NoStop}%
\bibitem [{\citenamefont {Thomas}\ \emph {et~al.}(2014)\citenamefont {Thomas},
  \citenamefont {Overy}, \citenamefont {Booth},\ and\ \citenamefont
  {Alavi}}]{thomas_symmetry_2014}%
  \BibitemOpen
  \bibfield  {author} {\bibinfo {author} {\bibfnamefont {R.~E.}\ \bibnamefont
  {Thomas}}, \bibinfo {author} {\bibfnamefont {C.}~\bibnamefont {Overy}},
  \bibinfo {author} {\bibfnamefont {G.~H.}\ \bibnamefont {Booth}}, \ and\
  \bibinfo {author} {\bibfnamefont {A.}~\bibnamefont {Alavi}},\ }\href
  {\doibase 10.1021/ct400835u} {\bibfield  {journal} {\bibinfo  {journal}
  {Journal of Chemical Theory and Computation}\ }\textbf {\bibinfo {volume}
  {10}},\ \bibinfo {pages} {1915} (\bibinfo {year} {2014})}\BibitemShut
  {NoStop}%
\bibitem [{\citenamefont {Petras}\ \emph {et~al.}(2019)\citenamefont {Petras},
  \citenamefont {Graham}, \citenamefont {Ramadugu}, \citenamefont
  {Goodpaster},\ and\ \citenamefont {Shepherd}}]{petras_fully_2019}%
  \BibitemOpen
  \bibfield  {author} {\bibinfo {author} {\bibfnamefont {H.~R.}\ \bibnamefont
  {Petras}}, \bibinfo {author} {\bibfnamefont {D.~S.}\ \bibnamefont {Graham}},
  \bibinfo {author} {\bibfnamefont {S.~K.}\ \bibnamefont {Ramadugu}}, \bibinfo
  {author} {\bibfnamefont {J.~D.}\ \bibnamefont {Goodpaster}}, \ and\ \bibinfo
  {author} {\bibfnamefont {J.~J.}\ \bibnamefont {Shepherd}},\ }\href {\doibase
  10.1021/acs.jctc.9b00571} {\bibfield  {journal} {\bibinfo  {journal} {Journal
  of Chemical Theory and Computation}\ }\textbf {\bibinfo {volume} {15}},\
  \bibinfo {pages} {5332} (\bibinfo {year} {2019})}\BibitemShut {NoStop}%
\bibitem [{\citenamefont {Rajagopal}\ \emph {et~al.}(1994)\citenamefont
  {Rajagopal}, \citenamefont {Needs}, \citenamefont {Kenny}, \citenamefont
  {Foulkes},\ and\ \citenamefont {James}}]{rajagopal_quantum_1994}%
  \BibitemOpen
  \bibfield  {author} {\bibinfo {author} {\bibfnamefont {G.}~\bibnamefont
  {Rajagopal}}, \bibinfo {author} {\bibfnamefont {R.~J.}\ \bibnamefont
  {Needs}}, \bibinfo {author} {\bibfnamefont {S.}~\bibnamefont {Kenny}},
  \bibinfo {author} {\bibfnamefont {W.~M.~C.}\ \bibnamefont {Foulkes}}, \ and\
  \bibinfo {author} {\bibfnamefont {A.}~\bibnamefont {James}},\ }\href
  {\doibase 10.1103/PhysRevLett.73.1959} {\bibfield  {journal} {\bibinfo
  {journal} {Physical Review Letters}\ }\textbf {\bibinfo {volume} {73}},\
  \bibinfo {pages} {1959} (\bibinfo {year} {1994})}\BibitemShut {NoStop}%
\bibitem [{\citenamefont {Shepherd}, \citenamefont {Henderson},\ and\
  \citenamefont
  {Scuseria}(2014{\natexlab{b}})}]{shepherd_range-separated_2014}%
  \BibitemOpen
  \bibfield  {author} {\bibinfo {author} {\bibfnamefont {J.~J.}\ \bibnamefont
  {Shepherd}}, \bibinfo {author} {\bibfnamefont {T.~M.}\ \bibnamefont
  {Henderson}}, \ and\ \bibinfo {author} {\bibfnamefont {G.~E.}\ \bibnamefont
  {Scuseria}},\ }\href {\doibase 10.1103/PhysRevLett.112.133002} {\bibfield
  {journal} {\bibinfo  {journal} {Physical Review Letters}\ }\textbf {\bibinfo
  {volume} {112}},\ \bibinfo {pages} {133002} (\bibinfo {year}
  {2014}{\natexlab{b}})}\BibitemShut {NoStop}%
\bibitem [{\citenamefont {Booth}, \citenamefont {Thom},\ and\ \citenamefont
  {Alavi}(2009)}]{booth_fermion_2009}%
  \BibitemOpen
  \bibfield  {author} {\bibinfo {author} {\bibfnamefont {G.~H.}\ \bibnamefont
  {Booth}}, \bibinfo {author} {\bibfnamefont {A.~J.~W.}\ \bibnamefont {Thom}},
  \ and\ \bibinfo {author} {\bibfnamefont {A.}~\bibnamefont {Alavi}},\ }\href
  {\doibase 10.1063/1.3193710} {\bibfield  {journal} {\bibinfo  {journal} {The
  Journal of Chemical Physics}\ }\textbf {\bibinfo {volume} {131}},\ \bibinfo
  {pages} {054106} (\bibinfo {year} {2009})}\BibitemShut {NoStop}%
\bibitem [{\citenamefont {Spencer}\ \emph {et~al.}(2019)\citenamefont
  {Spencer}, \citenamefont {Blunt}, \citenamefont {Choi}, \citenamefont
  {Etrych}, \citenamefont {Filip}, \citenamefont {Foulkes}, \citenamefont
  {Franklin}, \citenamefont {Handley}, \citenamefont {Malone}, \citenamefont
  {Neufeld}, \citenamefont {Di~Remigio}, \citenamefont {Rogers}, \citenamefont
  {Scott}, \citenamefont {Shepherd}, \citenamefont {Vigor}, \citenamefont
  {Weston}, \citenamefont {Xu},\ and\ \citenamefont
  {Thom}}]{spencer_hande-qmc_2019}%
  \BibitemOpen
  \bibfield  {author} {\bibinfo {author} {\bibfnamefont {J.~S.}\ \bibnamefont
  {Spencer}}, \bibinfo {author} {\bibfnamefont {N.~S.}\ \bibnamefont {Blunt}},
  \bibinfo {author} {\bibfnamefont {S.}~\bibnamefont {Choi}}, \bibinfo {author}
  {\bibfnamefont {J.}~\bibnamefont {Etrych}}, \bibinfo {author} {\bibfnamefont
  {M.-A.}\ \bibnamefont {Filip}}, \bibinfo {author} {\bibfnamefont {W.~M.~C.}\
  \bibnamefont {Foulkes}}, \bibinfo {author} {\bibfnamefont {R.~S.~T.}\
  \bibnamefont {Franklin}}, \bibinfo {author} {\bibfnamefont {W.~J.}\
  \bibnamefont {Handley}}, \bibinfo {author} {\bibfnamefont {F.~D.}\
  \bibnamefont {Malone}}, \bibinfo {author} {\bibfnamefont {V.~A.}\
  \bibnamefont {Neufeld}}, \bibinfo {author} {\bibfnamefont {R.}~\bibnamefont
  {Di~Remigio}}, \bibinfo {author} {\bibfnamefont {T.~W.}\ \bibnamefont
  {Rogers}}, \bibinfo {author} {\bibfnamefont {C.~J.~C.}\ \bibnamefont
  {Scott}}, \bibinfo {author} {\bibfnamefont {J.~J.}\ \bibnamefont {Shepherd}},
  \bibinfo {author} {\bibfnamefont {W.~A.}\ \bibnamefont {Vigor}}, \bibinfo
  {author} {\bibfnamefont {J.}~\bibnamefont {Weston}}, \bibinfo {author}
  {\bibfnamefont {R.}~\bibnamefont {Xu}}, \ and\ \bibinfo {author}
  {\bibfnamefont {A.~J.~W.}\ \bibnamefont {Thom}},\ }\href {\doibase
  10.1021/acs.jctc.8b01217} {\bibfield  {journal} {\bibinfo  {journal} {Journal
  of Chemical Theory and Computation}\ }\textbf {\bibinfo {volume} {15}},\
  \bibinfo {pages} {1728} (\bibinfo {year} {2019})}\BibitemShut {NoStop}%
\bibitem [{\citenamefont {Booth}\ \emph {et~al.}(2011)\citenamefont {Booth},
  \citenamefont {Cleland}, \citenamefont {Thom},\ and\ \citenamefont
  {Alavi}}]{booth_breaking_2011}%
  \BibitemOpen
  \bibfield  {author} {\bibinfo {author} {\bibfnamefont {G.~H.}\ \bibnamefont
  {Booth}}, \bibinfo {author} {\bibfnamefont {D.}~\bibnamefont {Cleland}},
  \bibinfo {author} {\bibfnamefont {A.~J.~W.}\ \bibnamefont {Thom}}, \ and\
  \bibinfo {author} {\bibfnamefont {A.}~\bibnamefont {Alavi}},\ }\href
  {\doibase 10.1063/1.3624383} {\bibfield  {journal} {\bibinfo  {journal} {The
  Journal of Chemical Physics}\ }\textbf {\bibinfo {volume} {135}},\ \bibinfo
  {pages} {084104} (\bibinfo {year} {2011})}\BibitemShut {NoStop}%
\bibitem [{\citenamefont {Booth}\ and\ \citenamefont
  {Alavi}(2010)}]{booth_approaching_2010}%
  \BibitemOpen
  \bibfield  {author} {\bibinfo {author} {\bibfnamefont {G.~H.}\ \bibnamefont
  {Booth}}\ and\ \bibinfo {author} {\bibfnamefont {A.}~\bibnamefont {Alavi}},\
  }\href {\doibase 10.1063/1.3407895} {\bibfield  {journal} {\bibinfo
  {journal} {The Journal of Chemical Physics}\ }\textbf {\bibinfo {volume}
  {132}},\ \bibinfo {pages} {174104} (\bibinfo {year} {2010})}\BibitemShut
  {NoStop}%
\bibitem [{\citenamefont {Shepherd}\ \emph
  {et~al.}(2012{\natexlab{b}})\citenamefont {Shepherd}, \citenamefont {Booth},
  \citenamefont {Grüneis},\ and\ \citenamefont {Alavi}}]{shepherd_full_2012}%
  \BibitemOpen
  \bibfield  {author} {\bibinfo {author} {\bibfnamefont {J.~J.}\ \bibnamefont
  {Shepherd}}, \bibinfo {author} {\bibfnamefont {G.}~\bibnamefont {Booth}},
  \bibinfo {author} {\bibfnamefont {A.}~\bibnamefont {Grüneis}}, \ and\
  \bibinfo {author} {\bibfnamefont {A.}~\bibnamefont {Alavi}},\ }\href
  {\doibase 10.1103/PhysRevB.85.081103} {\bibfield  {journal} {\bibinfo
  {journal} {Physical Review B}\ }\textbf {\bibinfo {volume} {85}},\ \bibinfo
  {pages} {081103} (\bibinfo {year} {2012}{\natexlab{b}})}\BibitemShut
  {NoStop}%
\bibitem [{Note1()}]{Note1}%
  \BibitemOpen
  \bibinfo {note} {With reference to the idea of a histogram}\BibitemShut
  {NoStop}%
\bibitem [{\citenamefont {Kwee}, \citenamefont {Zhang},\ and\ \citenamefont
  {Krakauer}(2008)}]{kwee_finite-size_2008}%
  \BibitemOpen
  \bibfield  {author} {\bibinfo {author} {\bibfnamefont {H.}~\bibnamefont
  {Kwee}}, \bibinfo {author} {\bibfnamefont {S.}~\bibnamefont {Zhang}}, \ and\
  \bibinfo {author} {\bibfnamefont {H.}~\bibnamefont {Krakauer}},\ }\href
  {\doibase 10.1103/PhysRevLett.100.126404} {\bibfield  {journal} {\bibinfo
  {journal} {Physical Review Letters}\ }\textbf {\bibinfo {volume} {100}},\
  \bibinfo {pages} {126404} (\bibinfo {year} {2008})}\BibitemShut {NoStop}%
\bibitem [{\citenamefont {Holzmann}\ \emph {et~al.}(2016)\citenamefont
  {Holzmann}, \citenamefont {Clay}, \citenamefont {Morales}, \citenamefont
  {Tubman}, \citenamefont {Ceperley},\ and\ \citenamefont
  {Pierleoni}}]{holzmann_theory_2016}%
  \BibitemOpen
  \bibfield  {author} {\bibinfo {author} {\bibfnamefont {M.}~\bibnamefont
  {Holzmann}}, \bibinfo {author} {\bibfnamefont {R.~C.}\ \bibnamefont {Clay}},
  \bibinfo {author} {\bibfnamefont {M.~A.}\ \bibnamefont {Morales}}, \bibinfo
  {author} {\bibfnamefont {N.~M.}\ \bibnamefont {Tubman}}, \bibinfo {author}
  {\bibfnamefont {D.~M.}\ \bibnamefont {Ceperley}}, \ and\ \bibinfo {author}
  {\bibfnamefont {C.}~\bibnamefont {Pierleoni}},\ }\href {\doibase
  10.1103/PhysRevB.94.035126} {\bibfield  {journal} {\bibinfo  {journal}
  {Physical Review B}\ }\textbf {\bibinfo {volume} {94}},\ \bibinfo {pages}
  {035126} (\bibinfo {year} {2016})}\BibitemShut {NoStop}%
\bibitem [{\citenamefont {Brown}\ \emph {et~al.}(2013)\citenamefont {Brown},
  \citenamefont {Clark}, \citenamefont {DuBois},\ and\ \citenamefont
  {Ceperley}}]{brown_path-integral_2013}%
  \BibitemOpen
  \bibfield  {author} {\bibinfo {author} {\bibfnamefont {E.~W.}\ \bibnamefont
  {Brown}}, \bibinfo {author} {\bibfnamefont {B.~K.}\ \bibnamefont {Clark}},
  \bibinfo {author} {\bibfnamefont {J.~L.}\ \bibnamefont {DuBois}}, \ and\
  \bibinfo {author} {\bibfnamefont {D.~M.}\ \bibnamefont {Ceperley}},\ }\href
  {\doibase 10.1103/PhysRevLett.110.146405} {\bibfield  {journal} {\bibinfo
  {journal} {Physical Review Letters}\ }\textbf {\bibinfo {volume} {110}},\
  \bibinfo {pages} {146405} (\bibinfo {year} {2013})}\BibitemShut {NoStop}%
\bibitem [{\citenamefont {Bulik}, \citenamefont {Chen},\ and\ \citenamefont
  {Scuseria}(2014)}]{bulik_electron_2014}%
  \BibitemOpen
  \bibfield  {author} {\bibinfo {author} {\bibfnamefont {I.~W.}\ \bibnamefont
  {Bulik}}, \bibinfo {author} {\bibfnamefont {W.}~\bibnamefont {Chen}}, \ and\
  \bibinfo {author} {\bibfnamefont {G.~E.}\ \bibnamefont {Scuseria}},\ }\href
  {http://arxiv.org/abs/1406.2034} {\bibfield  {journal} {\bibinfo  {journal}
  {arXiv:1406.2034 [physics]}\ } (\bibinfo {year} {2014})},\ \bibinfo {note}
  {arXiv: 1406.2034}\BibitemShut {NoStop}%
\bibitem [{\citenamefont {Knizia}\ and\ \citenamefont
  {Chan}(2012)}]{knizia_density_2012}%
  \BibitemOpen
  \bibfield  {author} {\bibinfo {author} {\bibfnamefont {G.}~\bibnamefont
  {Knizia}}\ and\ \bibinfo {author} {\bibfnamefont {G.~K.-L.}\ \bibnamefont
  {Chan}},\ }\href {\doibase 10.1103/PhysRevLett.109.186404} {\bibfield
  {journal} {\bibinfo  {journal} {Physical Review Letters}\ }\textbf {\bibinfo
  {volume} {109}},\ \bibinfo {pages} {186404} (\bibinfo {year}
  {2012})}\BibitemShut {NoStop}%
\bibitem [{\citenamefont {Zhu}, \citenamefont {Cui},\ and\ \citenamefont
  {Chan}(2020)}]{zhu_efficient_2020}%
  \BibitemOpen
  \bibfield  {author} {\bibinfo {author} {\bibfnamefont {T.}~\bibnamefont
  {Zhu}}, \bibinfo {author} {\bibfnamefont {Z.-H.}\ \bibnamefont {Cui}}, \ and\
  \bibinfo {author} {\bibfnamefont {G.~K.-L.}\ \bibnamefont {Chan}},\ }\href
  {\doibase 10.1021/acs.jctc.9b00934} {\bibfield  {journal} {\bibinfo
  {journal} {Journal of Chemical Theory and Computation}\ }\textbf {\bibinfo
  {volume} {16}},\ \bibinfo {pages} {141} (\bibinfo {year} {2020})},\ \bibinfo
  {note} {publisher: American Chemical Society}\BibitemShut {NoStop}%
\bibitem [{\citenamefont {Choi}\ \emph {et~al.}(2016)\citenamefont {Choi},
  \citenamefont {Kutepov}, \citenamefont {Haule}, \citenamefont {van
  Schilfgaarde},\ and\ \citenamefont {Kotliar}}]{choi_first-principles_2016}%
  \BibitemOpen
  \bibfield  {author} {\bibinfo {author} {\bibfnamefont {S.}~\bibnamefont
  {Choi}}, \bibinfo {author} {\bibfnamefont {A.}~\bibnamefont {Kutepov}},
  \bibinfo {author} {\bibfnamefont {K.}~\bibnamefont {Haule}}, \bibinfo
  {author} {\bibfnamefont {M.}~\bibnamefont {van Schilfgaarde}}, \ and\
  \bibinfo {author} {\bibfnamefont {G.}~\bibnamefont {Kotliar}},\ }\href
  {\doibase 10.1038/npjquantmats.2016.1} {\bibfield  {journal} {\bibinfo
  {journal} {npj Quantum Materials}\ }\textbf {\bibinfo {volume} {1}},\
  \bibinfo {pages} {1} (\bibinfo {year} {2016})},\ \bibinfo {note} {number: 1
  Publisher: Nature Publishing Group}\BibitemShut {NoStop}%
\bibitem [{\citenamefont {Zgid}\ and\ \citenamefont
  {Gull}(2017)}]{zgid_finite_2017}%
  \BibitemOpen
  \bibfield  {author} {\bibinfo {author} {\bibfnamefont {D.}~\bibnamefont
  {Zgid}}\ and\ \bibinfo {author} {\bibfnamefont {E.}~\bibnamefont {Gull}},\
  }\href {\doibase 10.1088/1367-2630/aa5d34} {\bibfield  {journal} {\bibinfo
  {journal} {New Journal of Physics}\ }\textbf {\bibinfo {volume} {19}},\
  \bibinfo {pages} {023047} (\bibinfo {year} {2017})}\BibitemShut {NoStop}%
\bibitem [{\citenamefont {Iskakov}\ \emph {et~al.}(2020)\citenamefont
  {Iskakov}, \citenamefont {Yeh}, \citenamefont {Gull},\ and\ \citenamefont
  {Zgid}}]{iskakov_ab_2020}%
  \BibitemOpen
  \bibfield  {author} {\bibinfo {author} {\bibfnamefont {S.}~\bibnamefont
  {Iskakov}}, \bibinfo {author} {\bibfnamefont {C.-N.}\ \bibnamefont {Yeh}},
  \bibinfo {author} {\bibfnamefont {E.}~\bibnamefont {Gull}}, \ and\ \bibinfo
  {author} {\bibfnamefont {D.}~\bibnamefont {Zgid}},\ }\href {\doibase
  10.1103/PhysRevB.102.085105} {\bibfield  {journal} {\bibinfo  {journal}
  {Physical Review B}\ }\textbf {\bibinfo {volume} {102}},\ \bibinfo {pages}
  {085105} (\bibinfo {year} {2020})},\ \bibinfo {note} {publisher: American
  Physical Society}\BibitemShut {NoStop}%
\bibitem [{\citenamefont {Goodpaster}\ \emph {et~al.}(2014)\citenamefont
  {Goodpaster}, \citenamefont {Barnes}, \citenamefont {Manby},\ and\
  \citenamefont {Miller}}]{goodpaster_accurate_2014}%
  \BibitemOpen
  \bibfield  {author} {\bibinfo {author} {\bibfnamefont {J.~D.}\ \bibnamefont
  {Goodpaster}}, \bibinfo {author} {\bibfnamefont {T.~A.}\ \bibnamefont
  {Barnes}}, \bibinfo {author} {\bibfnamefont {F.~R.}\ \bibnamefont {Manby}}, \
  and\ \bibinfo {author} {\bibfnamefont {T.~F.}\ \bibnamefont {Miller}},\
  }\href {\doibase 10.1063/1.4864040} {\bibfield  {journal} {\bibinfo
  {journal} {The Journal of Chemical Physics}\ }\textbf {\bibinfo {volume}
  {140}},\ \bibinfo {pages} {18A507} (\bibinfo {year} {2014})}\BibitemShut
  {NoStop}%
\bibitem [{\citenamefont {Spencer}\ \emph {et~al.}(2015)\citenamefont
  {Spencer}, \citenamefont {Blunt}, \citenamefont {Vigor}, \citenamefont
  {Malone}, \citenamefont {Foulkes}, \citenamefont {Shepherd},\ and\
  \citenamefont {Thom}}]{spencer_open-source_2015}%
  \BibitemOpen
  \bibfield  {author} {\bibinfo {author} {\bibfnamefont {J.~S.}\ \bibnamefont
  {Spencer}}, \bibinfo {author} {\bibfnamefont {N.~S.}\ \bibnamefont {Blunt}},
  \bibinfo {author} {\bibfnamefont {W.~A.}\ \bibnamefont {Vigor}}, \bibinfo
  {author} {\bibfnamefont {F.~D.}\ \bibnamefont {Malone}}, \bibinfo {author}
  {\bibfnamefont {W.~M.~C.}\ \bibnamefont {Foulkes}}, \bibinfo {author}
  {\bibfnamefont {J.~J.}\ \bibnamefont {Shepherd}}, \ and\ \bibinfo {author}
  {\bibfnamefont {A.~J.~W.}\ \bibnamefont {Thom}},\ }\href {\doibase
  10.5334/jors.bw} {\bibfield  {journal} {\bibinfo  {journal} {Journal of Open
  Research Software}\ }\textbf {\bibinfo {volume} {3}} (\bibinfo {year}
  {2015}),\ 10.5334/jors.bw}\BibitemShut {NoStop}%
\end{thebibliography}
 \end{document}